\documentclass[twocolumn,floatfix,showpacs,prd,aps,tightenlines,amssymb]{revtex4}
\usepackage{graphicx}
\usepackage{amsmath}
\usepackage{psfrag}
\usepackage{dcolumn}
\usepackage{bm}

\newcommand{\bea}{\begin{eqnarray}}
\newcommand{\eea}{\end{eqnarray}}
\newcommand{\beq}{\begin{equation}}
\newcommand{\eeq}{\end{equation}}
\newcommand{\KMS}{\rm km\,s^{-1}}

\newcommand{\SYLM}[3]{{Y^{#1}_{#2\, #3}}}
\newcommand{\SYLMP}[3]{{Y'^{#1}_{#2\, #3}}}

\begin{document}

\title{Comparison of Numerical and Post-Newtonian Waveforms for Generic
Precessing Black-Hole Binaries}

\author{Manuela Campanelli}
\author{Carlos O. Lousto}
\author{Hiroyuki Nakano}
\author{Yosef Zlochower} 
\affiliation{Center for Computational Relativity and Gravitation and
School of Mathematical Sciences,
Rochester Institute of Technology, 78 Lomb Memorial Drive, Rochester,
 New York 14623}

\date{\today}

\begin{abstract}
We compare waveforms and orbital dynamics from the first
long-term, fully non-linear, numerical simulations of a generic
black-hole binary configuration with post-Newtonian predictions. The
binary has mass ratio $q\sim0.8$ with arbitrarily oriented spins of
magnitude $S_1/m_1^2\sim0.6$ and $S_2/m_2^2\sim0.4$ and orbits 9 times
prior to merger. The numerical simulation starts with an initial separation of
$r\approx11M$, with orbital parameters determined by initial
 2.5PN and 3.5PN
post-Newtonian evolutions of a quasi-circular binary with an initial
separation of $r=50M$. The resulting binaries have very little
eccentricity according to the 2.5PN and 3.5PN systems, but show
significant eccentricities of $e\sim0.01-0.02$ and $e\sim0.002-0.005$
in the respective numerical simulations, thus demonstrating that 3.5PN
significantly reduces the eccentricity of the binary compared to
2.5PN. We perform three numerical evolutions from $r\approx11M$ with
maximum resolutions of $h=M/48,M/53.3,M/59.3$, to verify numerical
convergence. We observe a reasonably good agreement between the PN and
numerical waveforms, with an overlap of nearly 99\% for the first six
cycles of the $(\ell=2,m=\pm2)$ modes, 91\% for the $(\ell=2,m=\pm1)$
modes, and nearly 91\% for the $(\ell=3,m=\pm3)$ modes. The phase
differences between numerical and post-Newtonian approximations appear
to be independent of the $(\ell,m)$ modes considered and relatively
small for the first 3-4 orbits. An advantage of the 3.5 PN model over
the 2.5 PN one seems to be observed, which indicates that still higher
PN order (perhaps even 4.0PN) may yield significantly better
waveforms. In
addition, we identify features in the waveforms likely related to
precession and precession-induced eccentricity.

\end{abstract}

 \maketitle

\section{Introduction}
\label{sec:intro}

The discoveries of quasars, AGN, and other black-hole driven
astrophysical phenomena in the 1960's demonstrated that the most
energetic astrophysical phenomena are powered by gravity in the
strong-field regime. This, in turn, spurred a renewed interest in
classical General Relativity. The second major milestone in the revival
of the theory was the realization that when astrophysical black holes
merge, they release incredible amounts of energy in the
form of gravitational radiation, making them the brightest objects in the 
universe.  During their last few orbits, merging black-hole binaries 
release energy with a peak
luminosity of about $10^{-3}c^5/G$, $10^{23}$ times the power 
output of the Sun.

There are currently major experimental and theoretical efforts underway
to measure these gravitational wave signals. On the experimental side,
these efforts required the construction of kilometers long
interferometers, such as LIGO~\cite{LIGO3} and
VIRGO~\cite{Acernese:2004ru},
sensitive enough to measure arm length distance changes smaller than
the radius of a proton. While on the theoretical side, these efforts
required major advancements in signal extraction techniques and
the theoretical modeling of the gravitational wave sources. Modeling
the gravitational radiation from compact object sources has been
particularly difficult, as they require solving the fully non-linear
Einstein Equations of General Relativity on powerful supercomputers.
However, even with the rapid advancements in computer power,
solving the two-body problem in General Relativity proved to be remarkably difficult,
requiring over thirty years of research for the field to mature. Then in 2005, two
complementary and independent methods were discovered that allowed
numerical relativists to finally solve the black-hole binary
problem in full strong-field gravity~\cite{Pretorius:2005gq, Campanelli:2005dd, Baker:2005vv}.

The rapid progress and the number of new theoretical insights that
followed these breakthroughs have transformed the field of numerical
relativity (NR); turning it into a very valuable tool
with significant impact on
astrophysics~\cite{Campanelli:2004zw, Herrmann:2006ks, Baker:2006vn,
Sopuerta:2006wj, Gonzalez:2006md, Sopuerta:2006et, Herrmann:2006cd,
Herrmann:2007zz, Herrmann:2007ac, Campanelli:2007ew, Koppitz:2007ev,
Choi:2007eu, Gonzalez:2007hi, Baker:2007gi, Campanelli:2007cga,
Berti:2007fi, Tichy:2007hk, Herrmann:2007ex, Brugmann:2007zj,
Schnittman:2007ij, Krishnan:2007pu, HolleyBockelmann:2007eh,
Pollney:2007ss, Dain:2008ck, Redmount:1989, Merritt:2004xa,
Campanelli:2007ew, Gualandris:2007nm, HolleyBockelmann:2007eh,
Kapoor76, Bogdanovic:2007hp, Loeb:2007wz, Bonning:2007vt,
HolleyBockelmann:2007eh, Komossa:2008qd, Komossa:2008ye},
gravitational wave detection~\cite{Campanelli:2006gf, Baker:2006yw,
Campanelli:2006uy, Campanelli:2006fg, Campanelli:2006fy,
Pretorius:2006tp, Pretorius:2007jn, Baker:2006ha, Bruegmann:2006at,
Buonanno:2006ui, Baker:2006kr, Scheel:2006gg, Baker:2007fb,
Marronetti:2007ya, Pfeiffer:2007yz}, and on our theoretical
understanding of
black-binary spacetimes~\cite{Sperhake:2008ga, Hannam:2006vv,
Hannam:2008sg, Brown:2007tb, Garfinkle:2007yt, Dain:2008ck,
Krishnan:2007pu, Campanelli:2006fy, Campanelli:2008dv}.

One of the breakthrough methods, the `moving puncture'
approach~\cite{Campanelli:2005dd, Baker:2005vv}, was adopted by a
majority of the NR groups and has
proven to be accurate for the neutron-star binary and mixed
neutron-star---black-hole
binary problems~\cite{Etienne:2007jg, Yamamoto:2008js}, as well as for
black-hole configurations with more than two black holes~\cite{Campanelli:2007ea,
Lousto:2007rj}.

On the subject of black-hole binaries, the NR community is in 
very good agreement concerning a variety of results. 
Black-hole binaries will radiate between $2\%$ and $8\%$ of their
total mass and up to $40\%$ of their angular momenta, depending on the
magnitude and direction of the  spin components, during the
last few orbits and merger~\cite{Campanelli:2006uy, Campanelli:2006fg, Campanelli:2006fy,
Dain:2008ck}.
In general, these binaries will radiate net linear momentum, causing
the final remnant black hole to recoil~\cite{Campanelli:2004zw,  Herrmann:2006ks,
Baker:2006vn, Sopuerta:2006wj, Gonzalez:2006md, Sopuerta:2006et,
Herrmann:2006cd, Herrmann:2007zz, Herrmann:2007ac, Campanelli:2007ew,
Koppitz:2007ev, Choi:2007eu, Gonzalez:2007hi, Baker:2007gi,
Campanelli:2007cga, Berti:2007fi, Tichy:2007hk, Herrmann:2007ex,
Brugmann:2007zj, Schnittman:2007ij, Krishnan:2007pu,
HolleyBockelmann:2007eh, Pollney:2007ss, Dain:2008ck}. These recoils
can be very large when the black holes in the binary have significant
spin components in the orbital plane~\cite{Campanelli:2007ew,
Gonzalez:2007hi, Campanelli:2007cga, Healy:2008js} (up to $4000\ \KMS$
for astrophysical binaries~\cite{Campanelli:2007cga} and even $10000\
\KMS$ for extremely close hyperbolic encounters~\cite{Healy:2008js}),
which has astrophysically important effects~\cite{Redmount:1989,
Merritt:2004xa, Campanelli:2007ew, Gualandris:2007nm,
HolleyBockelmann:2007eh, Kapoor76}.  The observational consequences of
these large recoil velocities is an active area of current
research~\cite{Bogdanovic:2007hp, Loeb:2007wz, Bonning:2007vt,
HolleyBockelmann:2007eh, Komossa:2008qd, Komossa:2008ye}.

Currently, one of the most important tasks of NR is to
assist LIGO, VIRGO, and other
interferometric observatories, in detecting gravitational radiation
and extracting the  physical parameters of the sources. Given the demanding
resources required to generate these black-hole-binary simulations,
and the sheer volume of the seven-dimensional space of
intrinsic parameters of black-hole binaries, we need to develop
techniques to model arbitrary binary configuration based on numerical
simulations in a carefully chosen sample of the parameters space, in
combination with post-Newtonian and perturbative calculations.
One of the most promising of these approaches involves
determining the region of
common validity of the numerical simulations and post-Newtonian
expansions, with the goal of modeling the full waveform using
post-Newtonian waveforms for the initial inspiral and numerical
waveforms for the late-inspiral and merger. This method was pioneered
with the use of the Lazarus waveforms~\cite{Baker:2006ha} 
and has readily been pursued after the breakthroughs in NR. 

Comparisons of numerical simulations with post-Newtonian ones have
several benefits aside from the theoretical verification of PN.  From
a practical point of view, one can try to parametrize deviations of
the current 3.5PN expansions to fit the numerical
results~\cite{Pan:2007nw, Buonanno:2007pf, Damour:2007vq,
Damour:2008te, Boyle:2008ge}, or directly propose a phenomenological
description~\cite{Ajith:2007kx}, and thus make predictions in regions
of the parameter space still not explored by numerical simulations.
Another important application, from the theoretical point of view, is
to have a calibration of the post-Newtonian error in the last stages
of the binary merger.  The first results of comparisons for equal
mass, non-spinning binaries are encouraging~\cite{Buonanno:2006ui,
Baker:2006kr, Hannam:2007ik, Hinder:2008kv, Gualtieri:2008ux}.  Recently this analysis
was applied to equal-mass, equal-spin binaries with the spins aligned
with the orbital angular momentum (and thus
non-precessing)~\cite{Hannam:2007wf, Shoemaker:2008pe,
Vaishnav:2007nm}.

In this paper we compare the numerical and post-Newtonian
waveforms for the challenging problem of a generic black-hole binary,
i.e.\ a binary with unequal masses and  unequal, non-aligned, and
precessing spins.  The goal here is to evaluate accuracy of the
current order of post-Newtonian expansions when including spins
effects, as well as to develop new criteria for testing both numerical
and post-Newtonian developments.

The paper is organized as follows, in Sec.~\ref{sec:techniques}
we review the numerical techniques used for the evolution of
the black-hole binaries, in Sec.~\ref{sec:FN}
we present results from the numerical evolution of two similar generic
black-hole binaries, and in~\ref{sec:PN-intro} we analyze and compare 
different waveform modes as computed numerically and with the
highest available post-Newtonian approximation. Finally
in Sec.~\ref{sec:discussion} we present our conclusions.

\section{Techniques}
\label{sec:techniques}

To compute the numerical initial data, we use the puncture approach~\cite{Brandt97b}
along with the {\sc TwoPunctures}~\cite{Ansorg:2004ds} thorn.  In this
approach the 3-metric on the initial slice has the form $\gamma_{a b}
= (\psi_{BL} + u)^4 \delta_{a b}$, where $\psi_{BL}$ is the
Brill-Lindquist conformal factor, $\delta_{ab}$ is the Euclidean
metric, and $u$ is (at least) $C^2$ on the punctures.  The
Brill-Lindquist conformal factor is given by $ \psi_{BL} = 1 +
\sum_{i=1}^n m_{i}^p / (2 |\vec r - \vec r_i|), $ where $n$ is the
total number of `punctures', $m_{i}^p$ is the mass parameter of
puncture $i$ ($m_{i}^p$ is {\em not} the horizon mass associated with
puncture $i$), and $\vec r_i$ is the coordinate location of puncture
$i$.  We evolve these black-hole-binary data-sets using the {\sc
LazEv}~\cite{Zlochower:2005bj} implementation of the moving puncture
approach~\cite{Campanelli:2005dd,Baker:2005vv}.  In our version of the
moving puncture approach we replace the
BSSN~\cite{Nakamura87,Shibata95, Baumgarte99} conformal exponent
$\phi$, which has logarithmic singularities at the punctures, with the
initially $C^4$ field $\chi = \exp(-4\phi)$.  This new variable, along
with the other BSSN variables, will remain finite provided that one
uses a suitable choice for the gauge. An alternative approach uses
standard finite differencing of $\phi$~\cite{Baker:2005vv}.  Recently
Marronetti {\it et al.}~\cite{Marronetti:2007wz} proposed the use of
$W=\sqrt{\chi}$ as an evolution variable.  For the runs presented here
we use centered, eighth-order finite differencing in
space~\cite{Lousto:2007rj} and an RK4 time integrator (note that we do
not upwind the advection terms).

We use the Carpet~\cite{Schnetter-etal-03b} mesh refinement driver to
provide a `moving boxes' style mesh refinement. In this approach
refined grids of fixed size are arranged about the coordinate centers
of both holes.  The Carpet code then moves these fine grids about the
computational domain by following the trajectories of the two black
holes.

We obtain accurate, convergent waveforms and horizon parameters by
evolving this system in conjunction with a modified 1+log lapse and a
modified Gamma-driver shift
condition~\cite{Alcubierre02a,Campanelli:2005dd}, and an initial lapse
$\alpha(t=0) = 2/(1+\psi_{BL}^{4})$.  The lapse and shift are evolved
with
\begin{subequations}
\label{eq:gauge}
  \begin{eqnarray}
(\partial_t - \beta^i \partial_i) \alpha &=& - 2 \alpha K,\\
 \partial_t \beta^a &=& B^a, \\
 \partial_t B^a &=& 3/4 \partial_t \tilde \Gamma^a - \eta B^a.
 \label{eq:Bdot}
 \end{eqnarray}
 \end{subequations}
These gauge conditions require careful treatment of
$\chi$, the inverse of the three-metric conformal factor, near the
puncture in order for the system to remain
stable~\cite{Campanelli:2005dd,Campanelli:2006gf,Bruegmann:2006at}.
In practice one sets a floor value for $\chi$ equal to one-tenth of its initial
minimum value. This floor is only needed for the first $\sim5M$ of
evolution.
As shown in Ref.~\cite{Gundlach:2006tw},
this choice of gauge leads to a strongly hyperbolic
evolution system provided that the shift does not become too large.
In our tests, $W$ showed better behavior at very early times ($t <
10M$) (i.e.\ did not require any special treatment near the
punctures), but led to evolutions with lower effective resolution when
compared to $\chi$.
We chose $\eta=3$ for the simulations presented here.

We use {\sc AHFinderDirect}~\cite{Thornburg2003:AH-finding} to locate
apparent horizons.  We measure the magnitude of the horizon spin using
the Isolated Horizon algorithm detailed in~\cite{Dreyer02a}. This
algorithm is based on finding an approximate rotational Killing vector
(i.e.\ an approximate rotational symmetry) on the horizon $\varphi^a$. Given
this approximate Killing vector $\varphi^a$, the spin magnitude is
\begin{equation}
 \label{isolatedspin} S_{[\varphi]} =
 \frac{1}{8\pi}\int_{AH}(\varphi^aR^bK_{ab})d^2V,
\end{equation}
where $K_{ab}$ is the extrinsic curvature of the 3D-slice, $d^2V$ is
the natural volume element intrinsic to the horizon, and $R^a$ is the
outward pointing unit vector normal to the horizon on the 3D-slice.
We measure the direction of the spin by finding the coordinate line
joining the poles of this Killing vector field using the technique
introduced in~\cite{Campanelli:2006fy}.  Our algorithm for finding the
poles of the Killing vector field has an accuracy of $\sim 2^\circ$
(see~\cite{Campanelli:2006fy} for details). Note that once we have the
horizon spin, we can calculate the horizon mass via the Christodoulou
formula
\begin{equation}
{m^H} = \sqrt{m_{\rm irr}^2 +
 S^2/(4 m_{\rm irr}^2)},
\end{equation}
where $m_{\rm irr} = \sqrt{A/(16 \pi)}$ and $A$ is the surface area of
the horizon.

We also use an alternative quasi-local measurement of the spin and
linear momentum of the individual black holes in the binary that is
based on the coordinate rotation and translation
vectors~\cite{Krishnan:2007pu}.  In this approach the spin components
of the horizon are given by \begin{equation} S_{[i]} =
\frac{1}{8\pi}\int_{AH} \phi^a_{[i]} R^b K_{ab} d^2V,
\label{eq:coordspin} \end{equation} where $\phi^i_{[\ell]} =
\delta_{\ell j} \delta_{m k} r^m \epsilon^{i j k}$, and $r^m = x^m -
x_0^m$ is the coordinate displacement from the centroid of the hole,
while the linear momentum is given by \begin{equation} P_{[i]} =
\frac{1}{8\pi}\int_{AH} \xi^a_{[i]} R^b (K_{ab} - K \gamma_{ab})
d^2V, \label{eq:coordmom} \end{equation} where $\xi^i_{[\ell]} =
\delta^i_\ell$.

We measure radiated energy, linear momentum, and angular momentum, in
terms of $\psi_4$, using the formulae provided in
Refs.~\cite{Campanelli99,Lousto:2007mh}. However, rather than using
the full $\psi_4$, we decompose it into $\ell$ and $m$ modes and solve
for the radiated linear momentum, dropping terms with $\ell \geq 5$.
The formulae in Refs.~\cite{Campanelli99,Lousto:2007mh} are valid at
$r=\infty$. We obtain highly accurate values for these quantities by
solving for them on spheres of finite radius (typically $r/M=50, 60,
\cdots, 100$), fitting the results to a polynomial dependence in
$l=1/r$, and extrapolating to
$l=0$~\cite{Baker:2005vv,Campanelli:2006gf}. Each quantity $Q$ has the radial
dependence $Q=Q_0 + l Q_1 + {\cal O}(l^2)$, where $Q_0$ is the
asymptotic value (the ${\cal O}(l)$ error arises from the ${\cal
O}(l)$ error in $r\, \psi_4$). We perform both linear and quadratic
fits of $Q$ versus $l$, and take $Q_0$ from the quadratic fit as the
final value with the differences between the linear and extrapolated
$Q_0$ as a measure of the error in the extrapolations. We found that extrapolating
the waveform itself to $r=\infty$ introduced  phase errors due to
uncertainties in the areal radius of the observers, as well as
numerical noise. Thus when comparing PN to numerical waveforms, we use
the waveform extracted at $r=100M$. The extrapolations of the radiated
quantities are far more robust.

We convert the $(\ell,m)$ modes of $\psi_4$ into $(\ell,m)$ modes of
$h = h_{+} - i h_{\times}$ by calculating the Fourier transform
of each mode, dividing by $-\omega^2$ (where $\omega$ is the Fourier
frequency), setting the value of the
resulting transform to zero inside some specified window $-\omega_w <
\omega < \omega_w$, as well as chopping off the transform at
frequencies larger than 4 times the quasi-normal frequency,
 and finally taking the inverse transform.
By setting the transform to zero in this window, we remove the spurious
constant and linear terms from $h$ (we also remove spurious
high-frequency noise from the waveform by truncating the transform at
$\sim 4$ times the quasi-normal frequency). We confirm that the
calculation is correct by taking two time-derivatives of the resulting
$h$ and measuring how much the resulting function differs from
the original $\psi_4$ (See Fig.~\ref{fig:hdotdot_psi4} in
Sec.~\ref{sec:FN}).
We also use an alternative waveform comparison, based on the modes
of $\psi_4$ rather than $h$, which does not require this
transformation.

We compute the eccentricities of the orbits using the techniques
of~\cite{Husa:2007rh} and introduce a second technique based on
Newtonian trajectories. In~\cite{Husa:2007rh}, the eccentricity
$e_{D}$ is defined as
\begin{equation}
  \label{eq:ed}
  e_{D}(t) = \frac{r(t) - r_c(t)}{r_c(t)},
\end{equation}
where $r_c$ is obtained by fitting $r(t)$ to a low-order polynomial in
$t^{1/2}$. The actual eccentricity $e_{D}$ is the amplitude of the
oscillations in the resulting $e_{D}(t)$. We also introduce a second
measurement of eccentricity $e_{r}$ defined by
\begin{equation}
  \label{eq:er}
  e_{r}(t) = r(t)^2 \ddot r(t) / M.
\end{equation}
Here too, the eccentricity $e_r$ is the amplitude of the oscillations
in $e_r(t)$. This formula for the
eccentricity, which is only accurate for $e\ll1$, arises from the
Newtonian formula for the orbital radius 
$r(t) = \sqrt[3]{M/\Omega^2} (1+e \sin(\Omega t)) + {\cal O} (e^2)$.
Note that in both cases, $e(t)$ has sinusoidal oscillations and secular 
decay. The ellipticity is the amplitude of the sinusoidal
oscillations, while the secular decay affects the accuracy of the 
ellipticity calculation when its large. However, by differentiating
$r(t)$ twice with respect to $t$, the secular terms are suppressed.
Eq.~\ref{eq:er} can be modified with higher PN
corrections~\cite{Kidder:1995zr} to yield
\begin{equation}
\label{erddot}
e \cos(\Omega\,t)\approx\left[\ddot r(t)-\ddot r_0(t)\right]/(r\,\Omega^2),
\end{equation}
where
\begin{eqnarray}
\label{Omega}
\Omega^2=\frac{M}{r^3}\left[1-(3-\eta)(M/r)+{\cal O}(M/r)^2\right],\\
\dot r_0(t)=-\frac{64\eta}{5}\frac{M^3}{r^3}\left[1-\frac{1}{336}(1751+588\eta)(M/r)
\right],\\
\ddot
r_0(t)=\frac{16\eta}{105}\frac{M^5}{r^4}\left[252-(1751+588\eta)(M/r)\right],
\end{eqnarray}
and $r_0(t)$ is the zero-eccentricity inspiral trajectory.

\subsection{Initial Data}
\label{sec:ID}

To generate the initial data parameters, we used random values for the mass ratio
and spins of the binary (the ranges for these parameters were chosen
to make the evolution practical). We then calculated approximate
quasi-circular orbital parameters for a binary with these chosen
parameters at an initial orbital separation of $50M$ and evolved using
purely PN evolutions until the binary separation decreased to $11M$.
The goal was to produce very low
eccentricity orbital parameters at $r=11M$, as suggested
in~\cite{Husa:2007rh}. This technique is rather different from the
technique in~\cite{Pfeiffer:2007yz}, which used multiple numerical evolutions to
determine quasi-circular orbital parameters. The initial binary configuration
at $r=50M$ had $q=m_1/m_2 = 0.8$, $\vec S_1/m_1^2 = (-0.2, -0.14,
0.32)$, and $\vec S_2/m_2^2 =(-0.09, 0.48, 0.35)$. As described in
Sec.~\ref{sec:PN-intro}, we used both truncated 2.5PN equations of
motion for spinning binaries, and equations of motions including 
3.5PN corrections
(without the $H_{\rm S_1 S_2, 3PN}$ term). Our PN evolutions use the
ADM-TT gauge which is the one closest to the numerical quasi-isotropic
coordinates (to help reduce possible gauge
ambiguities)~\cite{Kelly:2007uc,Tichy:2002ec}.  We denote the two
resulting configuration by G2.5 and G3.5, respectively.  We then used
the PN momenta, spins, and particle locations to construct the initial
data for the numerical evolution.  We fixed the puncture masses by
requiring that the total ADM mass be $1M$ and that the mass ratio of
the two holes has the specified value. We renormalized the parameters
to obtain an ADM mass of $1M$ in order to aid comparison of the two
configurations and the analysis.

The initial data parameter are summarized in Table~\ref{table:ID}.
\begin{table}
\caption{Initial data parameters for the numerical evolutions.
Parameters for configuration G2.5 were obtained from a truncated
2.5PN evolution of a binary starting with an orbital separation of
$r=50M$, while parameters for configuration G3.5 were obtained from an
evolution with 3.5PN non-spinning corrections.  The punctures have
mass parameters $m^p_i$, horizons masses (Christodoulou) $m^H_i$,
momenta $\pm\vec p$, spins $\vec S_i$, and both configurations have a
total ADM mass $M_{\rm ADM}$. }
\label{table:ID}
\begin{ruledtabular}
\begin{tabular}{lccrcc}
            & G2.5 & G3.5 &             & G2.5 & G3.5 \\
\hline
$m^p_1/M$     & 0.40659 & 0.40694 & $m^p_2/M$     & 4.12328 & 0.456072 \\
$m^H_1/M$     & 0.44841 & 0.44833 & $m^H_2/M$     & 0.56054 & 0.56106 \\
$x_1/M$     & 3.32770 & -2.57272 & $x_2/M$     & -2.66216 & 2.05867 \\
$y_1/M$     & -5.15410 & -5.57057 & $y_2/M$     & 4.12328 & 4.45696 \\
$z_1/M$     & 0.51835 & -0.47758 & $z_2/M$     & -0.41468 & 0.40645 \\
$S^x_1/M^2$ & 0.017896 & -0.036840 & $S^x_2/M^2$ & -0.066727 & 0.025826 \\
$S^y_1/M^2$ & 0.069204 & -0.0050028 & $S^y_2/M^2$ & -0.098217 & 0.14951 \\
$S^z_1/M^2$ & 0.034786 & 0.069584 & $S^z_2/M^2$ &  0.14722 & 0.11050 \\
$p^x/M$     & 0.072919 & 0.080499 & $p^y/M$     & 0.048074 & -0.036311 \\
$10^{3} p^z/M$     & $-5.4117 $& $-0.743105$ & $M_{\rm ADM}/M$  & 1.00000  & 1.00000 \\
  
\end{tabular}
\end{ruledtabular}
\end{table}
We evolved these data using our eighth-order (in space) accurate code.
We evolved the G2.5 configuration using 12 levels of refinement,
with a finest
resolution of $h=M/48$, $M/53.33$, and $M/59.33$, and the outer
boundaries placed at $3072M$. We used the standard
5th-order Kreiss-Oliger dissipation operator and six buffer zones
at the refinement level boundaries. For the timestep, we chose a CFL
factor of $0.5$ for the inspiral phase, and then dropped the CFL by a
factor of $0.95$ during the merger phase. We reduced the CFL because
otherwise the
simulation proved to be unstable during the very fast plunge phase
(due to a violation of the CFL stability condition for our evolution
system).
We evolved the G3.5 configuration with the same setup as the $M/53.3$
G2.5 configuration, but chose an initial CFL factor of 0.475 (there
was no evidence of any instability with this reduced factor).

\section{Fully-nonlinear  numerical waveforms and trajectories}
\label{sec:FN}

\begin{figure}
\includegraphics[width=3.5in]{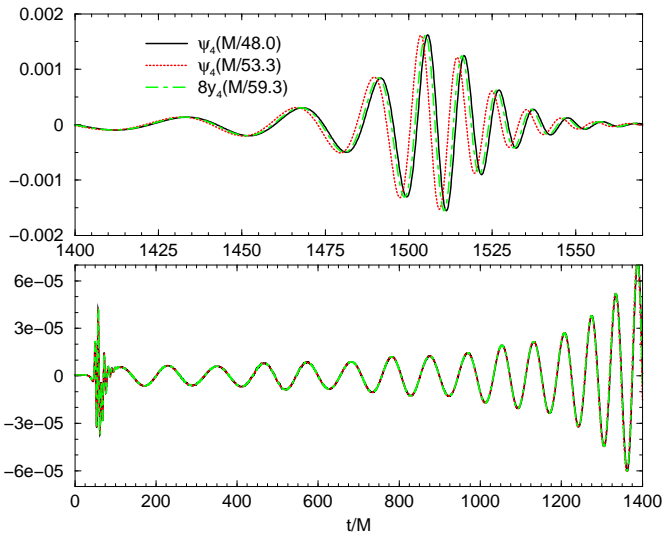}
\caption{The $(\ell=2, m=2)$ component of $\psi_4$ for the G2.5
configuration for the three
resolutions. Note the excellent phase agreement until about
$t=1400M$.}
\label{fig:NR_P4}
\end{figure}

\begin{figure}
\includegraphics[width=3.5in]{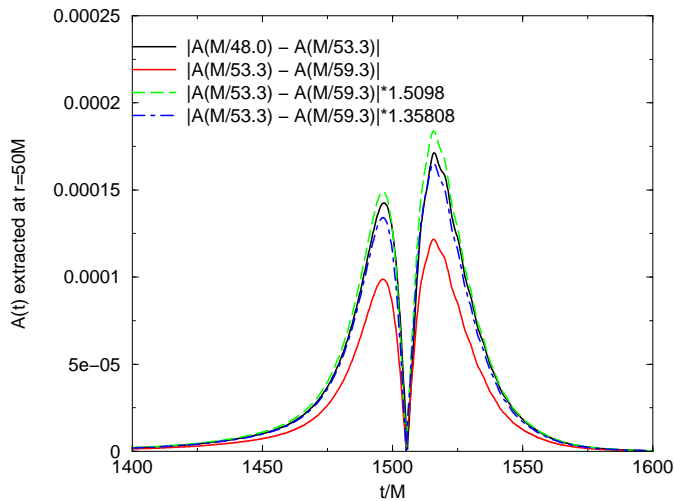}
\caption{Convergence of the amplitude of the G2.5 $(\ell=2,m=2)$ component
of $\psi_4$. The amplitude shows between 3rd- and 4th-order 
convergence (as demonstrated by multiplying the deviations in the
amplitude by 1.358 and 1.5098, respectively). }
\label{fig:NR_Amp_Conv}
\end{figure}

\begin{figure}
\includegraphics[width=3.5in]{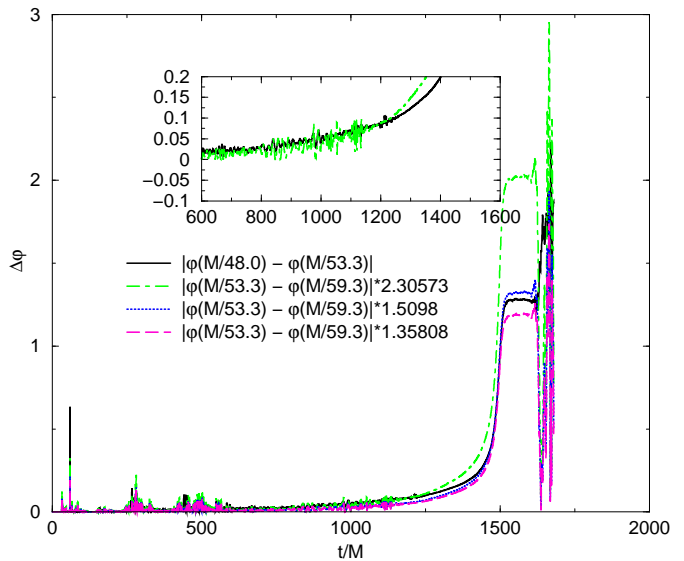}
\caption{Convergence of the G2.5 phase of the $(\ell=2,m=2)$ component
of $\psi_4$. The phase shows 8th-order convergence up to $t=1200M$,
decreasing to between 3rd- and 4th-order convergence during the plunge
(as demonstrated by multiplying the phase deviations by 2.305, 1.5098,
and 1.358 respectively). }
\label{fig:NR_Phase_Conv}
\end{figure}

\begin{figure}
\includegraphics[width=3.5in]{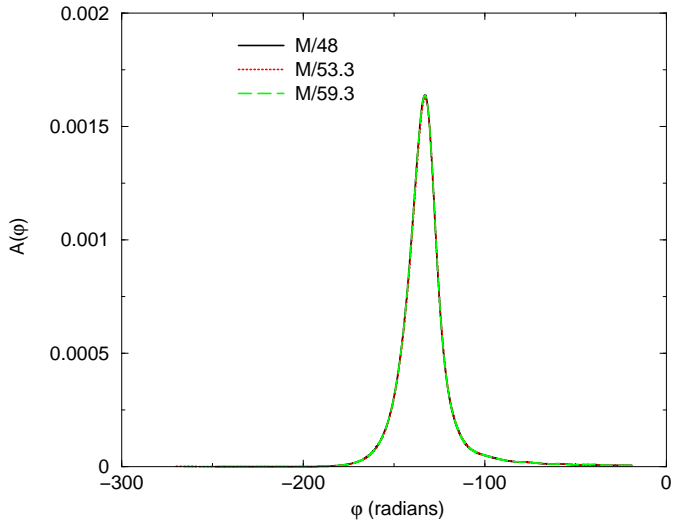}
\caption{The amplitude of the G2.5 $(\ell=2,m=2)$ component of $\psi_4$
versus the phase. Note that the phase becomes more negative as $t$
increases.}
\label{fig:NR_Amp_V_Phase}
\end{figure}

We calculated $\psi_4$ using our original 4th-order accurate
extraction code, and measured the convergence rate of the amplitude
and phase of the waveform separately. In Fig.~\ref{fig:NR_P4} we show
the $(\ell=2,m=2)$ component of $\psi_4$ of the G2.5 configuration
for the three resolutions.
Note the excellent phase agreement until about $t=1400M$. The phase
error increases exponentially during the last 2 orbits. In
Fig.~\ref{fig:NR_Amp_Conv} we show the convergence of the amplitude of
the $(\ell=2,m=2)$ mode. The amplitude shows between third- and
fourth-order convergence as is apparent by rescaling the amplitude
differences by $1.5098$ (fourth-order) and $1.35808$ (third-order).
As can be seen in Fig.~\ref{fig:NR_Phase_Conv}, the phase error
converges to eighth-order for $t < 1200M$. Beyond $t=1200M$ (which is
the beginning of the rapid plunge) the convergence falls to
fourth-order, as is apparent from the rescaling of the phase
differences by $2.30573$ (eighth-order), $1.5098$ (fourth-order), and
$1.35808$ (third-order). Note a convergence order of $8$ (up to
$t=1200M$) implies that the error in the phase for the highest
resolution run is less than $0.06$ radians for $t < 1200M$.
 In Fig.~\ref{fig:NR_Amp_V_Phase} we show the amplitude as a
function of phase. The phase error in the waveform converges to higher order than the
amplitude because it is sensitive to the phase error in the orbit,
which, in turn, is a function of the convergence of the evolution code.
The amplitude, however, appears to be sensitive to the extraction
algorithm's numerical error.

In Table~\ref{table:rad_results} we show the radiated energy, angular
momentum, and gravitational recoil versus resolution for the G2.5
configuration~\cite{Campanelli99,Lousto:2007mh}.
\begin{table}
  \caption{The radiated energy, angular
    momentum, and gravitational recoil versus resolution for the G2.5
configuration. The quoted uncertainties are due to extrapolation
$r\to\infty$. Note that this configuration has eccentricity
$e_{D}\sim0.02$ and $e_{r}\sim0.01$}
  \label{table:rad_results}
\begin{ruledtabular}
\begin{tabular}{lccc}
  & M/48 & M/53.3 & M/59.3 \\
\hline
$E_{\rm rad}/M$ & $0.0512\pm0.0039$ & $0.0513\pm0.0036$ & $0.0514\pm0.0033$ \\
$J^{x}_{\rm rad}/M^2$ & $0.018 \pm 0.021$ & $0.017 \pm 0.020$ & $0.014 \pm 0.013$ \\
$J^{y}_{\rm rad}/M^2$ & $-0.05 \pm 0.12$ & $-0.05 \pm 0.12$ & $-0.05 \pm 0.13$ \\
$J^{z}_{\rm rad}/M^2$ & $0.4445 \pm 0.0081$ & $0.4478 \pm 0.0103$ & $0.4466 \pm 0.0077$ \\
$V^{x}_{\rm rec} (\KMS)$ & $-1.6 \pm 5.7$ & $-6.9 \pm 6.0$ & $-2.2 \pm
5.5$ \\
$V^{y}_{\rm rec} (\KMS)$ & $78.36 \pm 6.51$ & $75.75 \pm 2.95$ &
$71.47 \pm 0.24$ \\
$V^{z}_{\rm rec} (\KMS)$ & $934  \pm 31$ & $1008 \pm 24$ & $947  \pm
16$ \\
\end{tabular}
\end{ruledtabular}
\end{table}
Here, extrapolation errors (to infinite radius) in the radiated energy and 
angular momenta dominate
the finite-difference errors, while the extrapolation errors in the
recoil appear to be similar to the finite difference errors. In
particular $V^z_{\rm rec}$ has a noticeable finite-difference error. This can be
understood in terms of the sensitivity of the out-of-plane recoil to
the angle that the spin direction makes with the infall direction at
merger. Thus orbital phase errors in the plunge can lead to significant
deviations in the out-of-plane recoil~\cite{Lousto:2007db,
Lousto:2008dn}.

The radiated energy, angular momentum, and the recoil
velocity for the G3.5 configuration are given in
Table~\ref{table:rad_G3.5results}.
The radiated energy and angular momenta are slightly larger for the
G3.5 configuration than the G2.5 configuration. Note that for both configurations, the radiated
angular momenta in the $x$ and $y$ directions are too small
to accurately measure. It should be pointed out that the quoted 
uncertainties
in the radiated quantities for G3.5 are due to extrapolation to
infinity. Additional uncertainties, due to truncation-errors are not
included (although results from G2.5 indicate that the uncertainties
in the radiated energy and angular momentum due to truncation errors
are small compared to the errors due to extrapolation). 
\begin{table}
  \caption{The radiated energy, angular
    momentum, and gravitational recoil for the G3.5
configuration. The quoted uncertainties are due to extrapolation
$r\to\infty$. Note that this configuration has eccentricity
$e_{D}\sim0.005$ and $e_r\sim0.002$}
  \label{table:rad_G3.5results}
\begin{ruledtabular}
\begin{tabular}{lc}
$E_{\rm rad}/M$ & $0.0522\pm0.0042$ \\
$J^{x}_{\rm rad}/M^2$ & $-0.20\pm0.27$\\
$J^{y}_{\rm rad}/M^2$ & $0.051\pm0.057$\\
$J^{z}_{\rm rad}/M^2$ & $0.4551\pm0.0029$\\
$V^{x}_{\rm rec} (\KMS)$ & $26.3\pm5.2$\\
$V^{y}_{\rm rec} (\KMS)$ & $103.0\pm5.7$\\
$V^{z}_{\rm rec} (\KMS)$ & $1529.9\pm8.9$\\
\end{tabular}
\end{ruledtabular}
\end{table}

\begin{figure}
\includegraphics[width=3.5in]{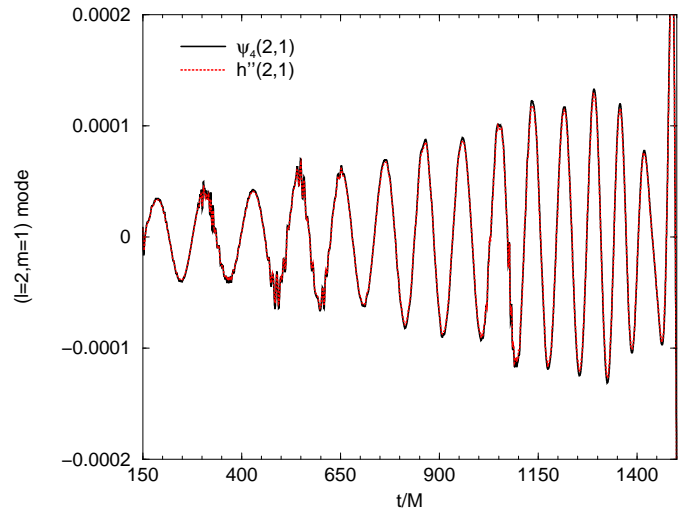}
\caption{A comparison of $\ddot h$ and $\psi_4$ for the $(l=2,m=1)$
mode for the G2.5 configuration. The plot demonstrates that the
windowing procedure apparently does not contaminate the waveform 
to a significant degree.}
\label{fig:hdotdot_psi4}
\end{figure}

As a final point, we show that our method for calculating $h$ from
$\psi_4$ using truncated Fourier transforms, yields a reasonable
approximation to the  original
$\psi_4$ after differentiating twice. In Fig.~\ref{fig:hdotdot_psi4}
we show $\ddot h$ and $\psi_4$ of the sub-leading $(l=2,m=1)$ mode of the 
G2.5 configuration (See however the discussion concerning the
amplitudes of $h$ in Sec.~\ref{sec:pn_amp}).

\subsection{Eccentricity and Precession}
\label{sec:num_ecc_and_pre}
\begin{figure}
\includegraphics[width=3.5in]{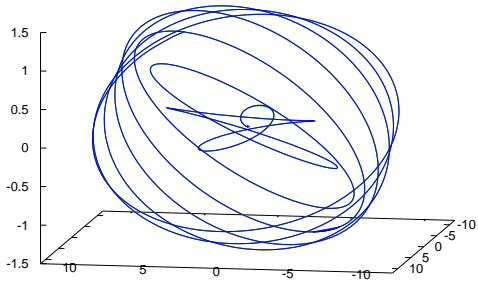}
\caption{The trajectory difference $\vec x_1 - \vec x_2$ for the G2.5
configuration. Note the
orbital plane precession and the very good agreement between trajectories
at the different resolutions (the tracks from the different resolutions
are not distinguishable on this scale).}
\label{fig:NR_3dtrack}
\end{figure}

\begin{figure}
\includegraphics[width=3.5in,height=3.43in]{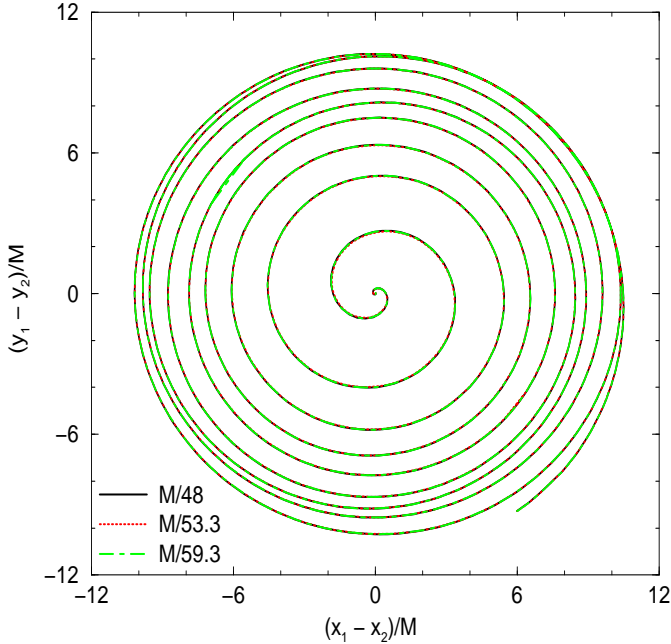}
\caption{An $xy$ projection of the trajectory difference $\vec x_1 -
\vec x_2$ for the G2.5 configuration. Note the very good agreement between trajectories
at the different resolutions. The initial orbital plane is inclined
with respect to the $xy$ plane, making the orbit appear more
eccentric.}
\label{fig:NR_2dtrack}
\end{figure}

\begin{figure}
\includegraphics[width=3.5in,height=3.43in]{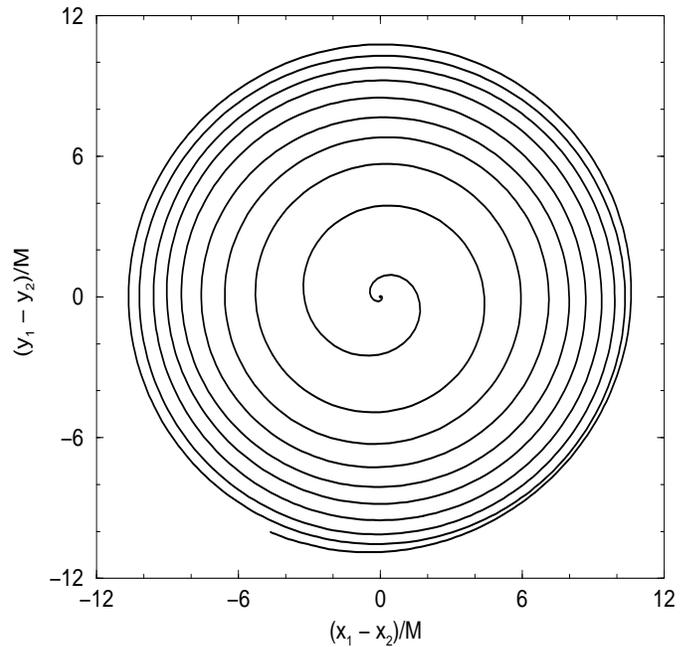}
\caption{An $xy$ projection of the trajectory difference $\vec x_1 -
\vec x_2$ for the G3.5 configuration.
The initial orbital plane is inclined
with respect to the $xy$ plane, making the orbit appear more
eccentric.}
\label{fig:NR_G3.5_2dtrack}
\end{figure}

\begin{figure}
\includegraphics[width=3.5in]{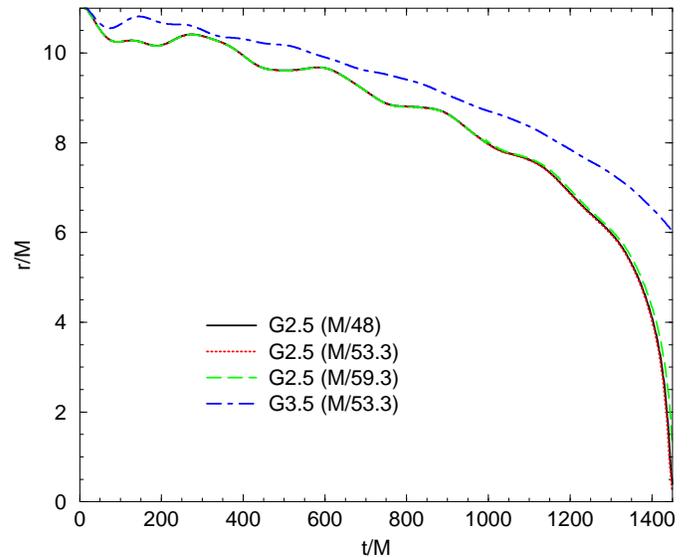}
\caption{The coordinate distance $r = |\vec x_1 - \vec x_2|$ between
punctures versus time for the G2.5 and G3.5 configurations. 
Note that the large eccentricity in the orbit
(apparent in the oscillation in $r$) is reduced by using the 3.5PN
equations to generate the initial data. Unlike in
Figs.~\ref{fig:NR_3dtrack} and~\ref{fig:NR_2dtrack}, here the
differences between resolution becomes apparent during the plunge.
These differences drive the phase error. Also note that the G3.5
configuration merges more slowly.}
\label{fig:NR_r_v_t}
\end{figure}

\begin{figure}
\includegraphics[width=3.5in]{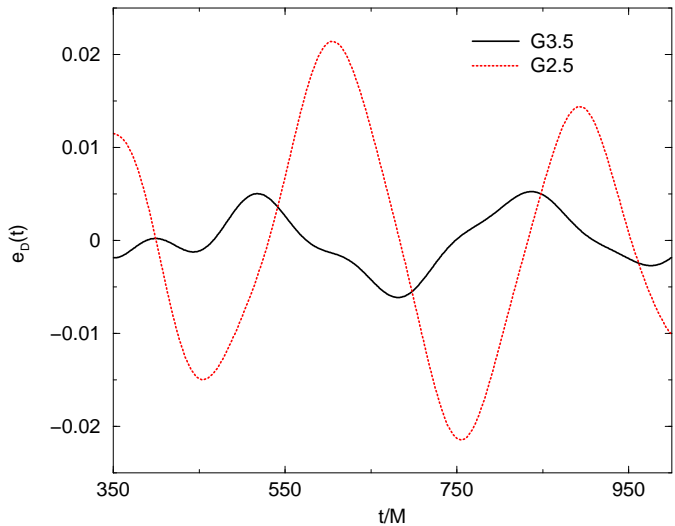}
\caption{The eccentricity $e_{D}(t)$ of the G3.5 and G2.5
configurations, as calculated using the techniques
of~\cite{Husa:2007rh}}
\label{fig:eccen}
\end{figure}

\begin{figure}
\includegraphics[width=3.5in]{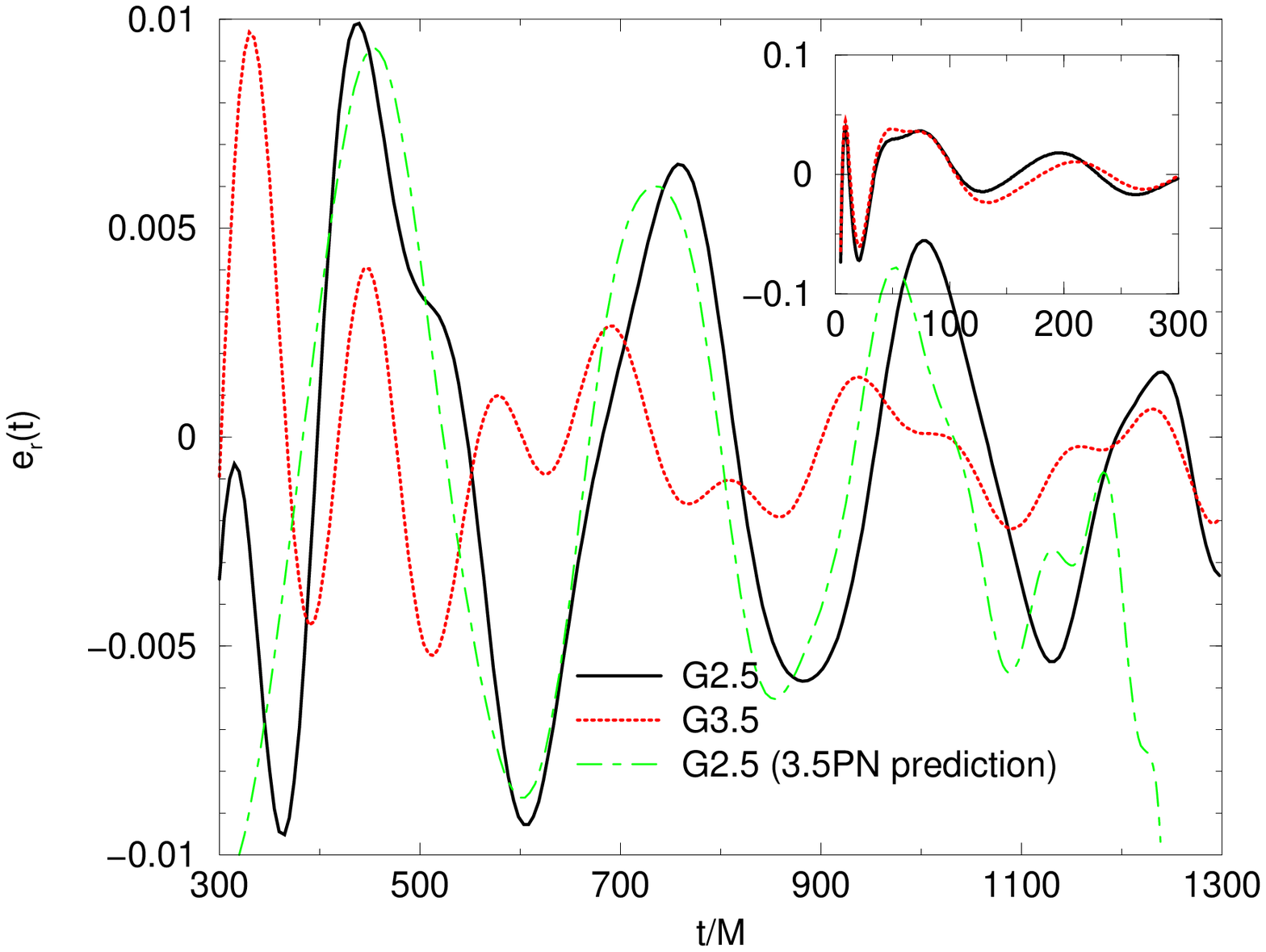}
\caption{The eccentricity $e_{r}(t)$ of the G3.5 and G2.5
configurations and the 3.5PN prediction for the G2.5 configuration
(the 3.5PN prediction for G3.5 is a factor of 10 smaller than the
NR prediction).
 The inset shows the `eccentricity' at early times when
gauge effects dominate the trajectories. Note that the eccentricity of
G2.5 decays throughout the evolution while the smaller eccentricity
for G3.5 remains roughly constant beyond $t\sim630M$. At later times the
eccentricities of G2.5 and G3.5 begin to agree.}
\label{fig:r_eccen}
\end{figure}

In Figs.~\ref{fig:NR_3dtrack},~\ref{fig:NR_2dtrack},
and~\ref{fig:NR_G3.5_2dtrack}, we show the orbital trajectory for the
G2.5 and G3.5 configurations. With the time
direction suppressed, we see excellent agreement between the
trajectories at the three resolutions. This is similar to the
excellent agreement in the amplitude versus phase of the $(\ell=2,
m=2)$ mode. However, when including time, as can be seen in
Fig.~\ref{fig:NR_r_v_t}, there is a significant difference between the
high and medium resolutions for $t>1200M$. Also note in
Fig.~\ref{fig:NR_r_v_t} the large eccentricity (apparent from the
oscillations in $r$) for the G2.5 configuration and that G3.5 has
reduced, but still large, eccentricity. Thus, assuming that the PN
series converges, we need to include still
higher-order PN correction to obtain low-eccentricity initial data
parameters. The reduced eccentricity of G3.5 compared to G2.5,
lends support to the hope that higher PN order will give low
eccentricity data. Alternatively, to produce low-eccentricity data,
one can try to use the iterative methods
of~\cite{Pfeiffer:2007yz}, which have been shown to work well for
non-spinning binaries. Using the methods of~\cite{Husa:2007rh}, we can
calculate the eccentricity $e_D(t)$, as shown in Fig.~\ref{fig:eccen}.
From the figure, we can see that the eccentricity of G2.5, which is
$e_{D}\sim 0.02$, is more than
3.5 times as large as the eccentricity of G3.5, which is $\sim0.005$.
Using the formula $e_{r}$ for the eccentricities (See
Fig.~\ref{fig:r_eccen}) yields $e_r\sim0.0088$ for G2.5 and
$e_r\sim0.0022$ for G3.5. However, as can be seen in the figure,
the eccentricity for G2.5 decays throughout the evolution, while the
eccentricity of G3.5 (although smaller than G2.5) remains roughly
constant for $t\gtrsim600M$. This is consistent with the results seen
in Fig.~\ref{fig:pn_ecc_gen} which shows that the 3.5 PN prediction
for the eccentricity does not decay with time for sufficiently close
binaries and small eccentricities.
In Ref.~\cite{Husa:2007rh}, they found that using PN parameters from
a PN-evolved inspiral (from $r=40M$ to $r=11M$) reduced the eccentricity
of the resulting binary from $e=0.01$, for a quasi-circular binary at
$r=11M$, to $e=0.002$. Here we
see eccentricities after a PN-evolved inspiral to $r=11M$ 
between 2.5 and 10 times as big.

\begin{figure}
\includegraphics[width=3.5in]{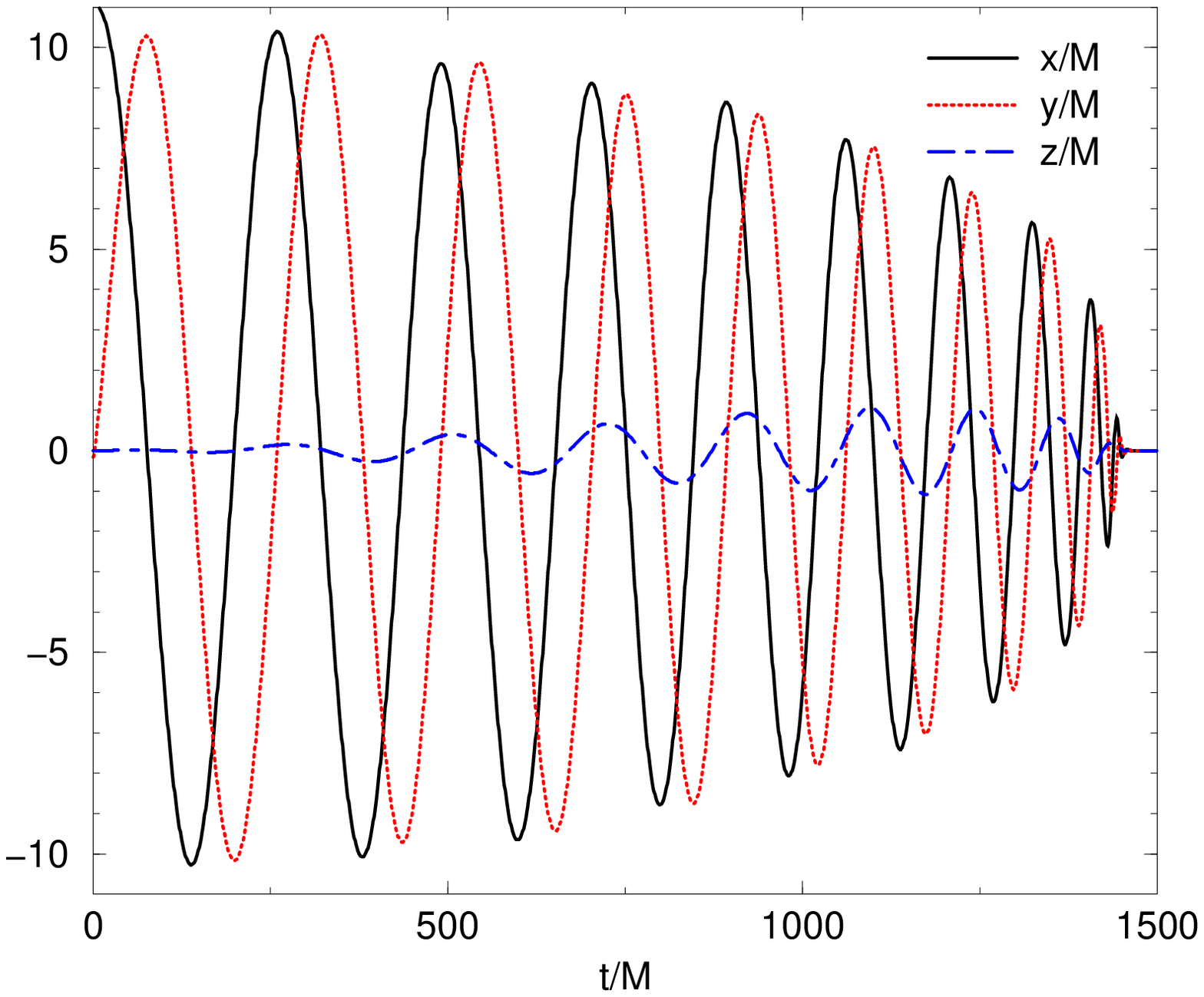}
\caption{The coordinate displacement $\vec r = \vec x_1 - \vec x_2$ between
punctures versus time for the G2.5 configuration after performing a
constant rotation that maps the initial orbital plane onto the $xy$
plane. Precession is responsible for driving the amplitude of $r^z$. }
\label{fig:G2.5_rotate}
\end{figure}

\begin{figure}
\includegraphics[width=3.5in]{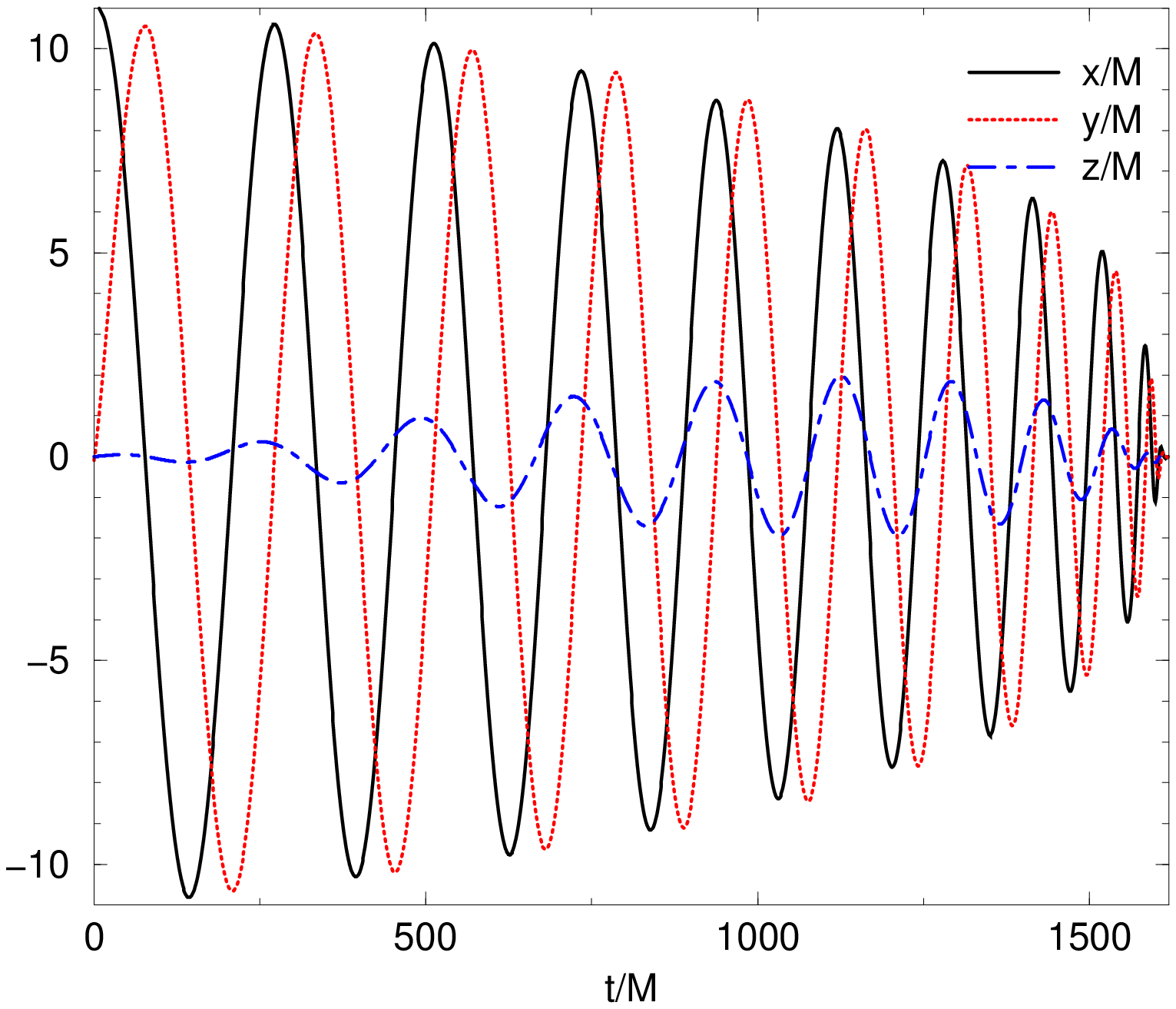}
\caption{The coordinate displacement $\vec r = \vec x_1 - \vec x_2$ between
punctures versus time for the G3.5 configuration after performing a
constant rotation that maps the initial orbital plane onto the $xy$
plane. Precession is responsible for driving the amplitude of $r^z$. }
\label{fig:G3.5_rotate}
\end{figure}

\begin{figure}
\includegraphics[width=3.5in]{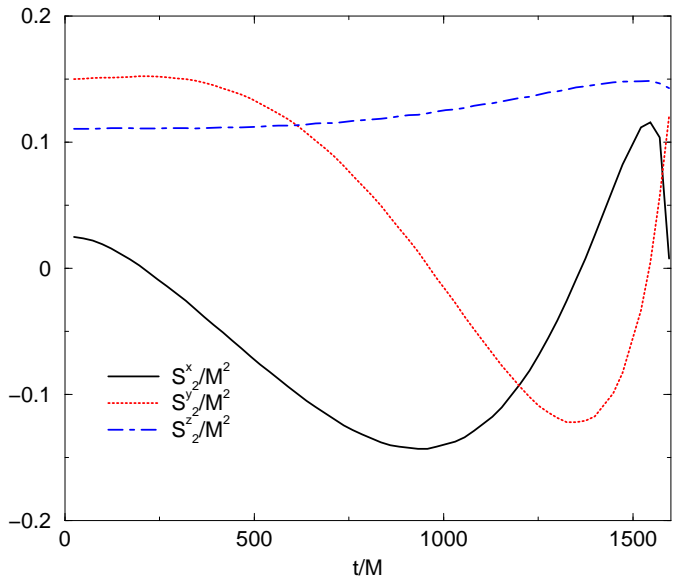}
\caption{The components of the spin for the more massive black hole in configuration
G3.5 as a function of time. The precession of the spin drive the orbital
plane precession. Here the precession timescale is of order $1000M$.}
\label{fig:G3.5_Precess}
\end{figure}

In Figs.~\ref{fig:G2.5_rotate} and~\ref{fig:G3.5_rotate} we show
$\vec r = \vec x_1 - \vec x_2$ versus time for the G2.5 and G3.5
configurations after performing a constant rotation that maps the
initial orbital motion onto the $xy$ plane. Orbital plane precession
drives the increase in amplitude of the $z$-component of $\vec r$.
The precession of the orbital plane is itself driven by the precession of the
total spin of the binary. Thus we can measure the rate of orbital plane
precession by looking at the components of the black-hole spins as a function
of time. In Fig.~\ref{fig:G3.5_Precess} we show the components of the
spin of the larger black hole as a function of time for the G3.5 configuration.
Note that the precessional frequency is quite low, with the precession
occurring on a timescale of order $1000M$; consistent with the time
scale in the amplitude modulation of the rotated $z_1 - z_2$
trajectory component in Figs.~\ref{fig:G2.5_rotate}~and~\ref{fig:G3.5_rotate}.
Despite this long timescale, precession can affect the waveform modes on
shorter timescales via mode-mixing effects. That is, precession of the
orbital plane will cause our mode decomposition (which uses a fixed
$z$-axis) to mix different modes (which oscillate at different frequencies).
This can lead to a beating effect that produces amplitude oscillations
visible in the waveform. In particular, when the orbital plane is
aligned with the $xy$ axis, the $m$ modes have a
frequency of $\sim m \omega_{\rm orbit}$. Hence if the $m=2$ and $m=1$
or $m=3$ modes mix, the
resulting system will have a beat frequency of $\sim \omega_{\rm orbit}$;
the same frequency as that due to eccentricity. Thus, oscillations in
the amplitude of the modes at the orbital frequency can arise both
from precession and ellipticity. We will come back to this point in
Sec.~\ref{sec:compare}.

\section{Post-Newtonian Equations of Motion and Waveforms}
\label{sec:PN-intro}

In order to calculate post-Newtonian (PN) waveforms, we need
to calculate the orbital motion of the binaries. We use the ADM-TT
gauge, which is the closest to our quasi-isotropic numerical initial
data coordinates~\cite{Kelly:2007uc,Tichy:2002ec}.  In this paper, we use two
different approximate PN equations of motion (EOM) based
on~\cite{Buonanno:2005xu, Damour:2007nc, Steinhoff:2007mb}.
To construct the EOM we use the Hamiltonian provided
in~\cite{Buonanno:2005xu}, with the additional terms provided
in~\cite{Damour:2007nc, Steinhoff:2007mb}, and the radiation-reaction
force provided in~\cite{Buonanno:2005xu}. We then use the standard
techniques of the Hamiltonian formulation to construct EOM for the
particle locations, momenta, and spins.
  In the
first approximate EOM, we included the purely orbital Hamiltonian up to
2PN order, spin-orbit coupling up to 2.5PN order, and spin-spin
coupling up to 2PN order (for the conservative part). That is to say,
we use the Hamiltonian
\begin{eqnarray}
H^{R} &=& H_{\rm O,Newt} + H_{\rm O,1PN} + H_{\rm O,2PN} 
\nonumber \\ &&
+ H_{\rm SO,1.5PN} + H_{\rm SO,2.5PN} + H_{\rm SS,2PN} \,.
\label{eq:H^R}
\end{eqnarray}
Here we only include the leading order radiation reaction
(dissipative) effect.  We refer to the above EOM as the ``truncated''
2.5PN EOM because there are terms up to 2.5PN order.  For the second
approximate EOM, we included the 3PN orbital Hamiltonian and 3PN
spin(1)-spin(2) coupling in the ADM-TT gauge~\cite{Steinhoff:2007mb}, i.e., we use the
Hamiltonian 
\begin{eqnarray}
H^{F} = H^{R} + H_{\rm O,3PN} + H_{\rm S_1S_2,3PN} \,
\label{eq:H^F}
\end{eqnarray} 
(the $H_{\rm S_1S_2,3PN}$ term was also computed in~\cite{Porto:2006bt,
 Porto:2007tt, Porto:2008tb} in a different gauge). 
For the dissipative part, 
we added the 3.5PN (non-spinning) radiation reaction terms, as well as the 
leading spin-orbit and spin-spin coupling 
to the radiation reaction~\cite{Buonanno:2005xu}. 
In the radiation reaction terms, 
we use the Taylor series of the
flux~\cite{Mroue:2008fu, Buonanno:2002ft}.
We refer to this second EOM  as 
the 3.5PN EOM (in practice the 3.5PN radiation reaction 
terms contribute to the orbital EOM at 6PN order).

We then use the following procedure to construct hybrid waveforms from
the orbital motion.  First we use the 1PN accurate waveforms derived
by Wagoner and Will~\cite{Wagoner:1976am} (WW waveforms) for a generic
orbit. By using these
waveforms, we can introduce effects due to eccentricity and effects
due to black-hole spins,
including the precession of the orbital plane.  On the other hand,
Blanchet {\it et al}.~\cite{Blanchet:2008je} recently obtained the
 3PN waveforms (B waveforms) for
non-spinning circular orbits.  We combine these two waveforms to
produce a hybrid waveform that includes the known higher-order
corrections to the waveform.  Note that, in the comparisons mentioned
below, the `truncated 2.5PN' waveforms and the 3.5 PN waveforms were
constructed from the same WW and B expressions. Differences only arise
because the `truncated' 2.5 PN waveforms are based on particle
trajectories obtained from the `truncated' 2.5 PN EOM.

 In order to combine the WW and B waveforms, we need to take into
account differences in the definitions of polarization states and the
angular coordinates (See Eqs.~(73)-(75) of~\cite{Wagoner:1976am} for
the definition of the WW polarization states
and Sec.~8 of~\cite{Blanchet:2008je} for the
definition of the B polarization states). 
The WW waveforms use the 
standard definition of GW polarization states,
which are the same as those derived from the Weyl scalar, but 
the B waveforms use an alternate definition; leading to a difference
in sign for all the $(\ell, m)$ modes of $h$.
The angular coordinates in the
B waveforms in~\cite{Blanchet:2008je} 
are derived from circular orbits in the equatorial ($xy$) plane. 
To directly compare the NR and PN waveforms, 
we must add an inclination to the B waveforms because in the generic
case the orbital
planes are inclined (with a time dependent inclination angle)
with respect to the $xy$ plane. Hence we need to
use the procedure developed in~\cite{Goldberg:1967, Gualtieri:2008ux}
to transform the $(\ell,m)$ modes
of  B waveforms into modes with respect to our rotated spin basis
(We provide a simple derivation of these transformations in
Appendix~\ref{app:proof}).
The following is an outline of the procedure. Let $\vec L = \vec r
\times \vec p$  be the instantaneous orbital
angular momentum, where
\begin{eqnarray}
\vec L &=& L (\sin \Theta_L \cos \Phi_L, \sin \Theta_L
 \sin \Phi_L, \cos \Theta_L),\\
\vec r &=& r (\sin \Theta_r \cos \Phi_r,
 \sin\Theta_r \cos \Phi_r, \cos \Theta_r ),\\
\vec p &=& p (\sin \Theta_p \cos \Phi_p, \sin\Theta_p \cos
\Phi_p, \cos \Theta_p ),
\end{eqnarray}
and $L$, $r$, $p$, $\Theta_L$, $\Phi_L$, $\Theta_r$, $\Phi_r$, $\Theta_p$,
$\Phi_p$ are functions of time.
 The first step is to rotate the
orbital plane onto the $xy$ plane. Let ${\bf R}(\alpha, \beta, \gamma)$ be a
general rotation defined by the Euler angles $\alpha$,
$\beta$, and $\gamma$, where we first perform a rotation through angle
$\alpha$ about the $z$ axis, followed by a rotation through angle
$\beta$ about the $y$ axis, and finally a rotation through angle
$\gamma$ about the $z$ axis (in practice, we never need to perform
this final rotation). Thus a rotation ${\bf R}(-\Phi_L(t), -\Theta_L(t), 0)$
transforms $\vec L$ and $\vec r$ into $\vec L'$ and $\vec r'$, where
\begin{eqnarray}
\vec L'&=& L(0,0,1),\\
{\vec r}'&=& r(\cos \Phi_B(t), \sin \Phi_B (t), 0).
\end{eqnarray}
The $(\ell,m)$ modes of
the B waveform, in a frame where the orbital plane is
the $xy$ plane, can be written in terms of $\cos \Phi_B(t)$, $\sin
\Phi_B(t)$, $r$, and $\omega_{\rm orbit}$.
In order to calculate the $(\ell,m)$ modes of $h$ with respect to the
numerical coordinates (where the orbital plane is inclined), we use
the results of~\cite{Goldberg:1967, Gualtieri:2008ux}.
As was shown in~\cite{Goldberg:1967}, the spin-weighted $s$ spherical
harmonics in the numerical coordinates are related to those in the
rotated coordinates (where the orbital plane is the $xy$ plane) by
\begin{equation}
  \SYLM{s}{\ell}{m}(\Omega) = e^{i s \chi} \sum_{m'} e^{-i m' \alpha} K^{\ell\,
s}_{m' m}(-\beta) e^{-i m \gamma} \SYLM{s}{\ell}{m'}(\Omega'),
\end{equation}
where $\alpha$, $\beta$, and $\gamma$
 are the rotation angles described above (note $\gamma=0$),
and the phase factor $e^{i s \chi}$ arises from the transformation of
spin-weighted function under a change of spin-basis.
In~\cite{Gualtieri:2008ux} it was shown that $K^{\ell\, s}_{m' m}$ is
independent of $s$ (see Appendix~\ref{app:proof} for an alternative
proof), and is thus given by~\cite{Goldberg:1967}
\begin{equation}
  K^{\ell\, s}_{m' m}(-\beta) = d^{\ell}_{m' m} (-\beta),
\end{equation}
where $d^{\ell}_{m' m} (\beta)$ is the Wigner $d$ matrix
given by,
\begin{eqnarray}
  d^{\ell}_{m' m} (\beta) =
\sqrt{(\ell+m)!(\ell-m)!(\ell+m')!(\ell-m')!} \nonumber \\
\times \sum_k \frac{(-1)^{k+m'-m}}{k!(\ell+m-k)!(\ell-m'-k)!(m'-m+k)!}
\nonumber \\
\left(\sin \frac{\beta}{2} \right)^{2k+m'-m}
\left(\cos \frac{\beta}{2} \right)^{2\ell-2k-m'+m},
\end{eqnarray}
where the sum over $k$ is such that the factorials are non-negative.
Since $h = h' e^{-2 i \chi}$, we have
\begin{eqnarray}
h_{\ell m} &=& \int h \overline {\SYLM{-2}{\ell}{m}} d \Omega \nonumber \\
 &=& \sum_{m'}\int h' e^{i m' \alpha} d^{\ell}_{m' m}(-\beta) h'
\overline {\SYLM{-2}{\ell}{m'}} (\Omega') d\Omega' \nonumber \\
 &=& \sum_{m'}e^{i m' \alpha} d^{\ell}_{m' m}(-\beta)  h'_{\ell m'} \nonumber
\\
 &=& \sum_{m'}e^{-i m' \Phi_L} d^{\ell}_{m' m}(\Theta_L)  h'_{\ell m'}.
\end{eqnarray}

The remaining complication arises from the fact that both the WW and B
waveforms contain terms for a non-spinning circular orbit. To avoid
adding the common terms twice, we subtract them from the B waveforms.
First, using the 1PN WW formulae, we obtain the waveforms from
non-spinning circular orbits in the equatorial plane. We do this by
applying the 3PN EOM for circular orbits to the WW waveform formulae.
We then rewrite the waveforms in terms of the gauge invariant variable
$x$, which is the normalized frequency. The B waveforms are given in
terms of $x$, so we can identify those terms in the WW waveforms also
present in B waveforms in a unique way. We then remove these terms
from the B waveforms.
 For our generic
case, we rotate the subtracted B waveforms modes and add them to the
modes of the WW waveforms to obtain the hybrid waveform.
Note that there are no significant gauge ambiguities arising from
combining the WW and B waveforms in this way because at 1PN order the
harmonic and ADM gauges are equivalent (and hence the WW waveforms are
the same in the two gauges) and the B waveforms are given in terms of
gauge invariant variables.

Note that we calculate the spin contribution to the waveform
through its effect on the orbital motion directly in the WW waveforms
and indirectly in B waveforms through the inclination of the orbital
plane. Other effects of spin and orbital plane precession on the
waveforms are currently not known.

\subsection{Orbital motion and initial parameters}

Following the procedure detailed in~\cite{Husa:2007rh}, extended to
spinning particles, we used purely
post-Newtonian evolutions of a nearly quasi-circular binary with
initial orbital separation $r=50M$ to obtain the positions, momenta,
and spins for a non-eccentric binary with separation $r\sim 11M$.
The idea behind this procedure is that one can specify quasi-circular
parameters with very low eccentricity for binaries with large
separations using the conservative part of the Hamiltonian (i.e.\
solve for circular orbits). The
subsequent PN evolution then provides the PN parameters (including
radial momentum) of a closer binary with similar (but lower) eccentricity. 
The initial quasi-circular binary configuration
at $r=50M$
had PN parameters $q=m_1/m_2 = 0.8$, $\vec S_1/m_1^2 = (-0.2, -0.14, 0.32)$, and
$\vec S_2/m_2^2 =(-0.09, 0.48, 0.35)$.  We refer to the binary
configurations obtained using the truncated 2.5PN and 3.5PN EOM as
G2.5 and G3.5, respectively. It turns out that the order of the PN
evolution is critical for producing low eccentricity binaries. The
eccentricity of the G2.5 configuration, as measured by a subsequent
2.5PN evolution is quite small. However, both the 
numerical and 3.5PN simulations, show a that the
eccentricity for G2.5 is actually relatively large.
Similarly, the eccentricity of the
G3.5 configuration, as determined from the full numerical simulation,
 while smaller than the G2.5 configuration, is still relatively large.  We
used these $r\sim11M$ parameters
in our numerical and subsequent PN evolutions.

\begin{figure}
\includegraphics[width=3.5in]{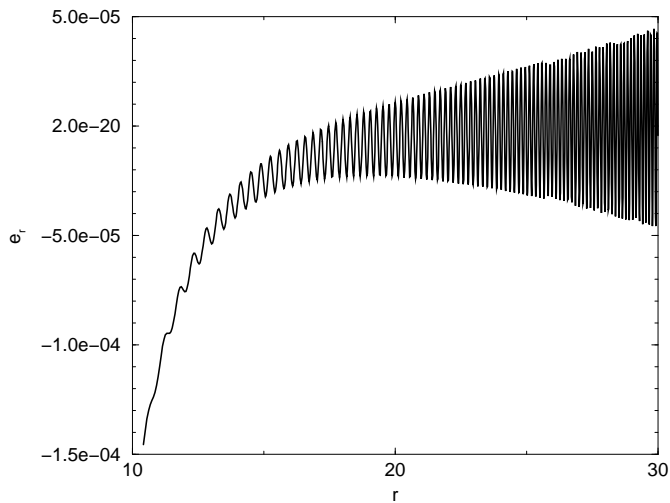}
\caption{$e_{r}$ versus radius for a binary with spins aligned with the
angular momentum. Here the eccentricity decreases with $r$ for all
radii.}
\label{fig:pn_ecc_up_up}
\end{figure}

\begin{figure}
\includegraphics[width=3.5in]{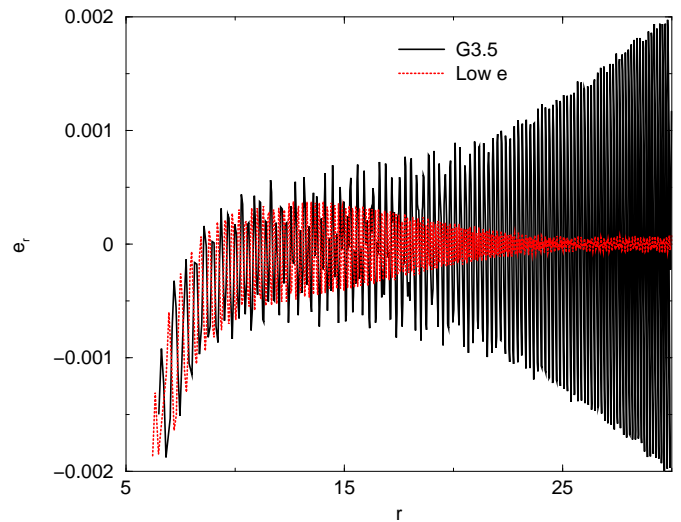}
\caption{$e_{r}$ versus radius for the G3.5 configuration and a very
similar binary, with parameters chosen to reduce the (PN) initial
eccentricity. Note that the eccentricity at $r<10$ is constant and
roughly the same for both configurations.}
\label{fig:pn_ecc_gen}
\end{figure}

It is interesting to note that in the generic case, the eccentricity,
according to 3.5PN does not decrease with time at smaller radii. To
demonstrate this, we show the eccentricity, calculated using the
formula $e_{r} (t) = r^2 \ddot r / M$, where the magnitude of the
oscillations in $e_{r}(t)$ is the eccentricity. In Fig.~\ref{fig:pn_ecc_up_up} we show the eccentricity versus time for
a configuration with the same spin-magnitudes and mass ratio as our
generic case, but with the spins aligned with the orbital angular
momentum. As can be seen, the eccentricity decreases with radius.
However, in Fig.~\ref{fig:pn_ecc_gen} we show the eccentricity
calculated for our configuration, and one slightly modified to give an
even lower initial eccentricity, versus time. Here we see that the
eccentricity decreases to about $e\sim0.0005$ and then remains
constant. On the other hand, for the low-eccentricity data, the
eccentricity actually increases until reaching $e\sim 0.0005$.
From the figures is apparent that precession affects induce an
apparent ellipticity to the binary's motion that is not radiated away
(at least to this order in the PN expansion).

\begin{table}
  \caption{PN orbital parameters for the G2.5 and G3.5 configuration
at an orbital separation of $r\sim11M$, as calculated directly from PN
simulations starting at $r=50M$. $m_1$ and $m_2$ denote the masses,
$x$, $y$, and $z$ denote the
components of ${\vec r} = {\vec x}_1 -{\vec x}_2$, $p_i$ ($i=x,y,z$)
denotes the linear momentum, and $S_{1i}$
and $S_{2i}$ denote the spin angular momenta.}
  \label{table:PN_11M_PARAMS}
\begin{ruledtabular}
\begin{tabular}{lcc}
 & G2.5 & G3.5\\
\hline
  $m_1/M$ &  0.4455115640 & 0.4455115640 \\
  $m_2/M$    &  0.5568894551 & 0.5568894551 \\
  $x/M$      &  5.9453450513 & -4.5976488271 \\
  $y/M$      & -9.2084320770 & -9.9544694746 \\
  $z/M$      &  0.9260944396 & -0.8775891873 \\
  $p_x/M$    &  0.0723766737 & 0.0799120544 \\
  $p_y/M$    &  0.0477169131 & -0.0360468994 \\
  $p_z/M$    & -0.0053715184 & -0.00073769138 \\
  $S_{1x}/M^2$ &  0.0176308357 & -0.0365711851 \\
  $S_{1y}/M^2$ &  0.0681788517 & -0.0049664012 \\
  $S_{1z}/M^2$ &  0.0342713607 & 0.0690768531 \\
  $S_{2x}/M^2$ & -0.0657393278 & 0.0256376428 \\
  $S_{2y}/M^2$ & -0.0967624976 & 0.1484228759 \\
  $S_{2z}/M^2$ &  0.1450366736 & 0.1096979400 
\end{tabular}
\end{ruledtabular}
\end{table}
For the G2.5 configuration we used a truncated 2.5PN evolution,
which began at $r=50M$, to
obtain the PN parameters provided in Table~\ref{table:PN_11M_PARAMS}.
The specific spins of the
two holes are $S_1/m_1^2= 0.3945883931$ and $S_2/m_2^2 =
0.6008327554$, respectively.  The 2.5PN ADM mass, $M_{\rm
ADM}=m_1+m_2+H^R$, for these parameters is $M_{\rm ADM}/M =
0.9925682736,$ where $H^R$ is given by Eq.~(\ref{eq:H^R}).

\begin{figure}[ht]
\includegraphics[width=3.5in]{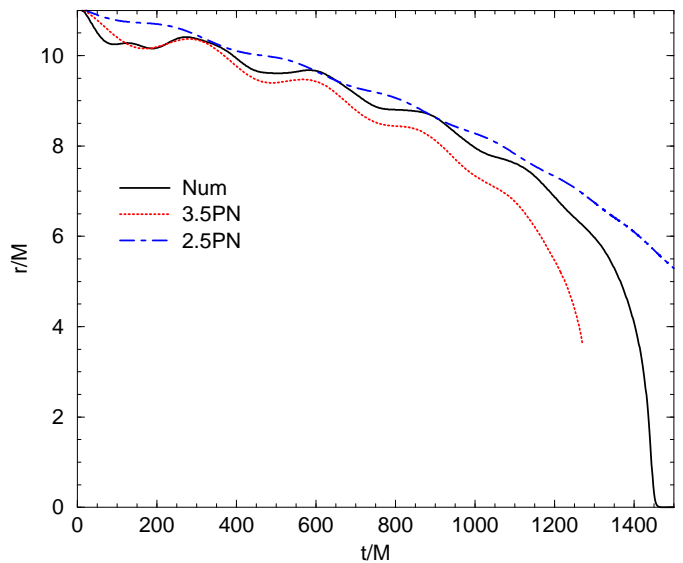}
\caption{The evolution of the orbital radius for the G2.5
configuration from the numerical, 2.5PN, and 3.5PN simulations.  The
residual eccentricity in the 2.5PN evolution is due to our using a
different 2.5PN radiation reaction term from that used in the original
evolution beginning at $r=50M$.  Note that both 3.5PN and the numerical
simulation indicate that this configuration has relatively large
eccentricity.
}
\label{fig:G2.5_PN_NUM_R}
\end{figure}

When using these parameters in the numerical evolution, and
subsequent PN evolutions starting from $r/M_{\rm ADM}=11.08236108$, we
normalized the PN parameters by the ADM Mass (i.e.\ we use the parameters $\vec r\to \vec
r/M_{\rm ADM}$, $\vec p \to \vec p/M_{\rm ADM}$, and $\vec S\to\vec
S/M_{\rm ADM}^2$). This renormalization is helpful because we choose
to normalize our numerical simulations such that the total ADM mass is
1.
However, due to the spurious radiation on the initial slice, the
numerical black-hole masses change with time, and eventually equilibrate to a
mass ratio of $q=0.7993$ (the uncertainty in the numerical masses of
the two holes was $\delta m\sim 0.00003$ at the highest resolutions).
 Thus in order to compare the PN and
numerical waveforms, we need to account for this change in mass ratio.
To do this, we modified our choices of $m_1$ and $m_2$ such that $M_{\rm
ADM}/M=1$ and $q=m_1/m_2=0.7993$.  However, because our two PN
evolutions systems have different Hamiltonians, we needed to use
slightly different values of $m_1/M_{\rm ADM}$ and $m_2/M_{\rm ADM}$
in each case. Note that the spin angular momentum is not affected by
the spurious radiation to a significant level because the spurious
radiation is nearly axially symmetric about the two holes.
For the truncated 2.5PN evolutions we used
\begin{eqnarray}
m_1/M_{\rm ADM} &=& 0.4486274928 \,, \nonumber \\
m_2/M_{\rm ADM} &=& 0.5612754821 \,,
\end{eqnarray}
i.e.\ from the equation $M_{\rm ADM}=1=(q+1)m_2 + H^R(q,m_2)$, 
while for the 3.5PN evolutions we used
\begin{eqnarray}
m_1/M_{\rm ADM} &=& 0.4486635058 \,, \nonumber \\
m_2/M_{\rm ADM} &=& 0.5613205377 \,,
\end{eqnarray}
i.e.\ $M_{\rm ADM}=1=(q+1)m_2 + H^F(q,m_2)$. 
We verified that these changes in the masses have a negligible effect
on the eccentricity and waveforms according to the PN evolutions.
We then used both the truncated 2.5PN and 3.5PN equations of motion to
evolve this modified configuration from $r\approx11M$.  We made one
additional change in the truncated 2.5PN evolution of G2.5. In our
original truncated 2.5PN evolution from $r=50M$, we used a simpler
form of the radiation reaction term based on PN expansion in the orbital parameters
$r$ and $\vec p$. While in the subsequent evolution, we used a new
expression (consistent with the old expression to 2.5PN order in the
Taylor expansion of the PN orbital parameters) based
on an expansion in the orbital frequency~\cite{Buonanno:2005xu}.
However, because we changed the EOM, the
truncated 2.5PN evolution of the G2.5 configuration, which according
to the original system  had very-low eccentricity, now has a
small residual eccentricity (see Fig.~\ref{fig:G2.5_PN_NUM_R}).
The radiation-reaction terms is directly related to the radial
motion of the binary. Therefore, the radiation reaction force
is very important to determine the quasi-circular configuration,
and differences in the force have a strong effect on the motion.
This is an indication that 2.5PN is not accurate enough to model
the binary's motion in the $r=50M$ to $r=11M$ range.

For the G3.5 configuration, we used a 3.5PN evolution (that did not
include the $H_{\rm S_1S_2,3PN}$ term) from
$r=50M$ to $r=11M$ to obtain the orbital parameters provided in
Table~\ref{table:PN_11M_PARAMS}.
The 3.5PN ADM mass of this system is $M_{\rm ADM}/M = 0.9927145092$,
and, once again, we renormalized the PN parameters by the ADM mass.
When evolving this system numerically, we used slightly altered
values of the spin
\begin{eqnarray}
  S_{1x} / M_{\rm ADM}^2 &=& -0.0368395795 \,, \nonumber \\ 
  S_{1y} / M_{\rm ADM}^2 &=& -0.0050028494 \,, \nonumber \\ 
  S_{1z} / M_{\rm ADM}^2 &=&  0.0695838052 \,, \nonumber \\ 
  S_{2x} / M_{\rm ADM}^2 &=&  0.0258257964 \,, \nonumber \\ 
  S_{2y} / M_{\rm ADM}^2 &=&  0.1495121453 \,, \nonumber \\ 
  S_{2z} / M_{\rm ADM}^2 &=&  0.1105030087 \,,
\end{eqnarray}
which introduced negligible changes in the waveforms and eccentricity.
Here too, we find that the black holes absorb spurious radiation arising from
the initial data that changes the mass ratio to $0.79937$. To model
this change in the 3.5PN evolution, we changed the $m_1$ and $m_2$ PN
masses to $m_1/M_{\rm ADM} =  0.4485829815$ and $m_2/M_{\rm ADM} =
0.5611706488$. Here too, the changes to the masses do not affect the motion
or eccentricity of the binary according to the 3.5PN evolution.
Thus, one should use an iterative procedure, like those in
Refs.~\cite{Boyle:2007ft, Baker:2006kr}, to reduce the eccentricity.

\begin{figure}[ht]
\includegraphics[width=3.5in]{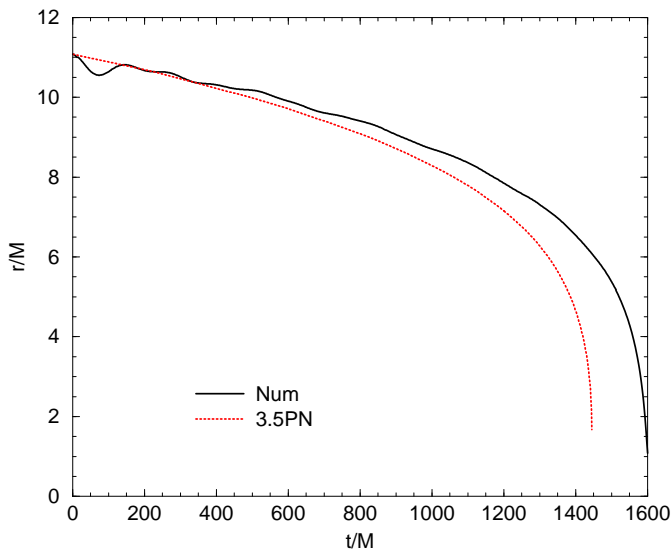}
\caption{The evolutions of the orbital radius for the G3.5
configuration from the numerical and 3.5PN simulations. Here the
numerical simulations shows that the
eccentricity was reduced, but is still relatively large,
while the 3.5PN evolution indicates that the binary is non-eccentric.}
\label{fig:G3.5_PN_NUM_R}
\end{figure}

According to the truncated 2.5PN evolution (with the new radiation
reaction term based on the orbital frequency discussed above),
the G2.5 configuration
has a relatively small eccentricity, as is apparent in the small
oscillations of the time
dependence of the 2.5 PN orbital radius displayed in
Fig.~\ref{fig:G2.5_PN_NUM_R}. However,
both a subsequent 3.5PN evolution and the numerical evolution
showed that these data were highly eccentric. In
Fig.~\ref{fig:G2.5_PN_NUM_R} we see that both the 3.5PN and numerical
simulations produce similar, large orbital radius oscillations (which
are due to eccentricity).  The G3.5 configurations, which has
very-low eccentricity according to 3.5PN, as is apparent in
the non-oscillatory behavior of the 3.5PN orbital radius seen in
Fig.~\ref{fig:G3.5_PN_NUM_R}, still shows relatively large
oscillations in the orbital radius of the numerical simulation.
Thus, using the 3.5PN equations of motion to generate low-eccentricity
initial data reduces the eccentricity, but not nearly to the extent
seen in non-spinning binaries~\cite{Husa:2007rh}.

\subsection{Comparison of NR and PN waveforms}
\label{sec:compare}

\begin{figure}[ht]
\includegraphics[width=3.5in]{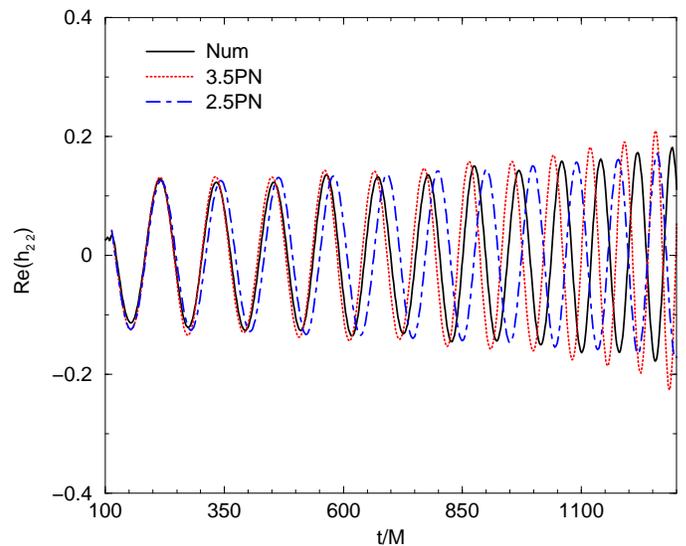}
\caption{The real part of the $(\ell=2, m=2)$ mode of $h$ for the G2.5
configuration from the numerical, truncated 2.5PN, and 3.5PN simulations. 
Note that the 3.5PN prediction is closer to the numerical waveform and
that 3.5PN predicts an early merger while 2.5PN predicts a late
merger (as is evident by the amplitude of the mode versus time).
 }
\label{fig:G2.5_re_h22}
\end{figure}

\begin{figure}
\includegraphics[width=3.5in]{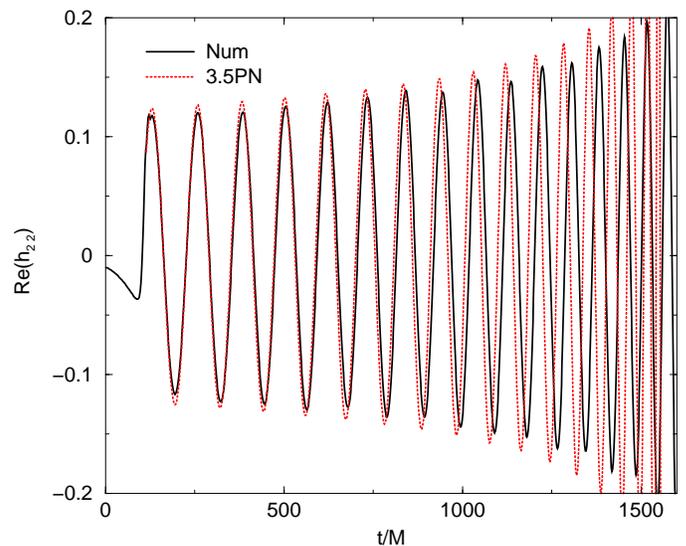}
\caption{The real part of the $(\ell=2, m=2)$ mode of $h$ for the G3.5
configuration from the numerical and 3.5PN simulations.
Here too, 3.5PN predicts an early merger (as is evident by the
amplitude of the mode versus time).
}
\label{fig:G3.5_re_h22}
\end{figure}

\begin{figure}[ht]
\includegraphics[width=3.5in]{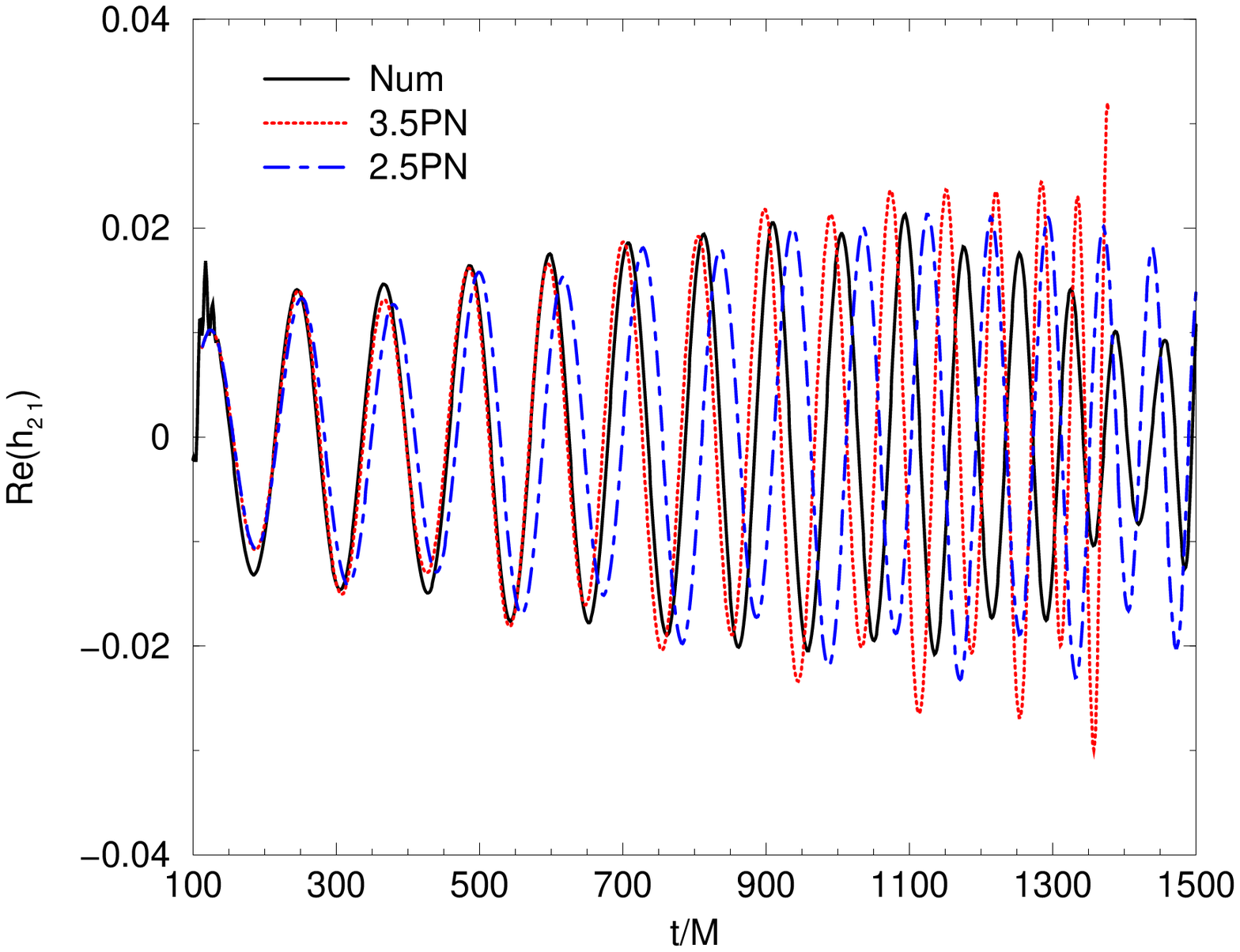}
\caption{
The real part of the $(\ell=2, m=1)$ mode of $h$ for the G2.5
configuration from the numerical, truncated 2.5PN, and 3.5PN simulations. 
Note the precession induced modulation in the amplitude of the
oscillations.
}
\label{fig:G2.5_re_h21}
\end{figure}

\begin{figure}
\includegraphics[width=3.5in]{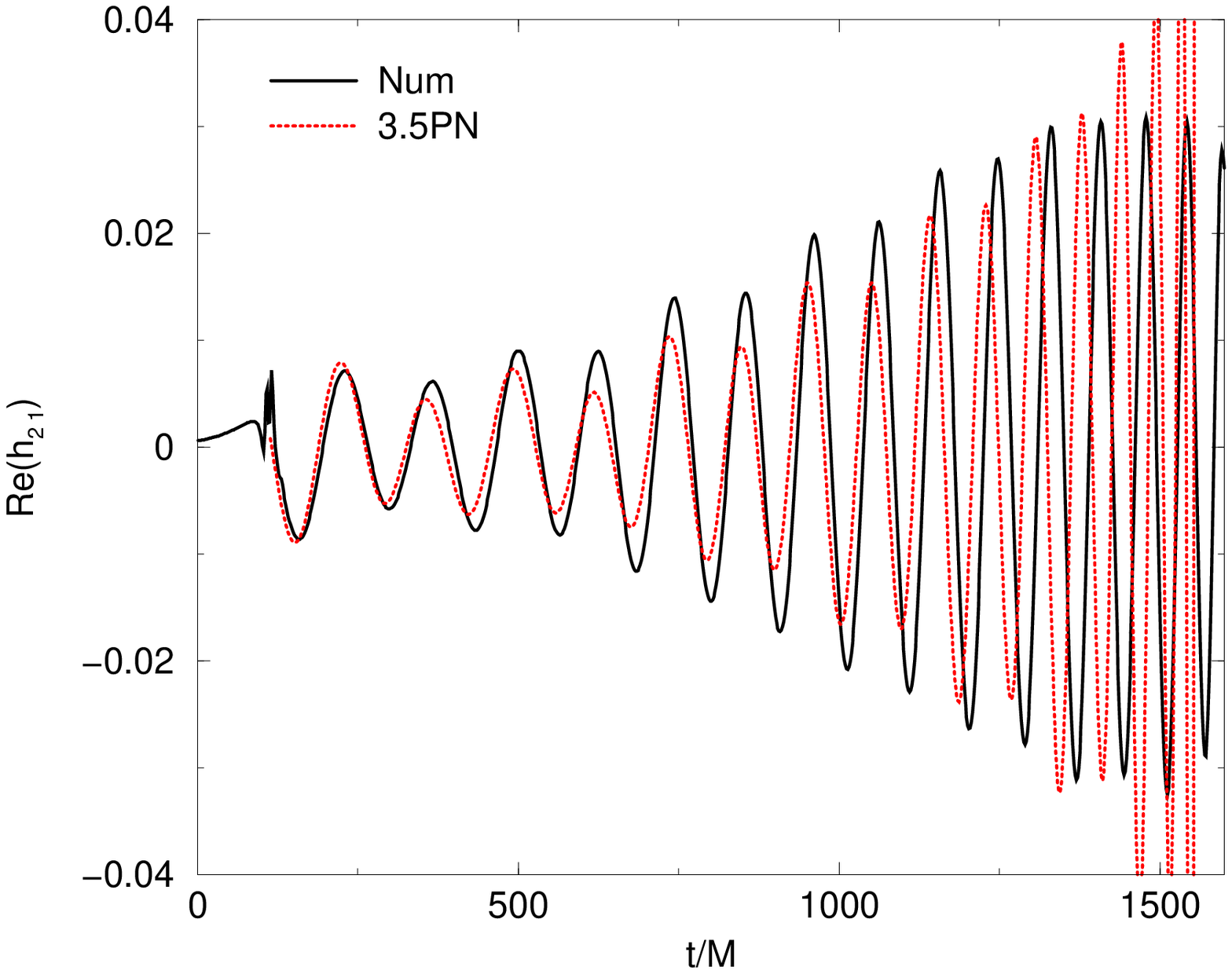}
\caption{The real part of the $(\ell=2, m=1)$ mode of $h$ for the G3.5
configuration from the numerical and 3.5PN simulations. Note the
precession induced modulation in the amplitude of the
oscillations.
}
\label{fig:G3.5_re_h21}
\end{figure}

\begin{figure}[ht]
\includegraphics[width=3.5in]{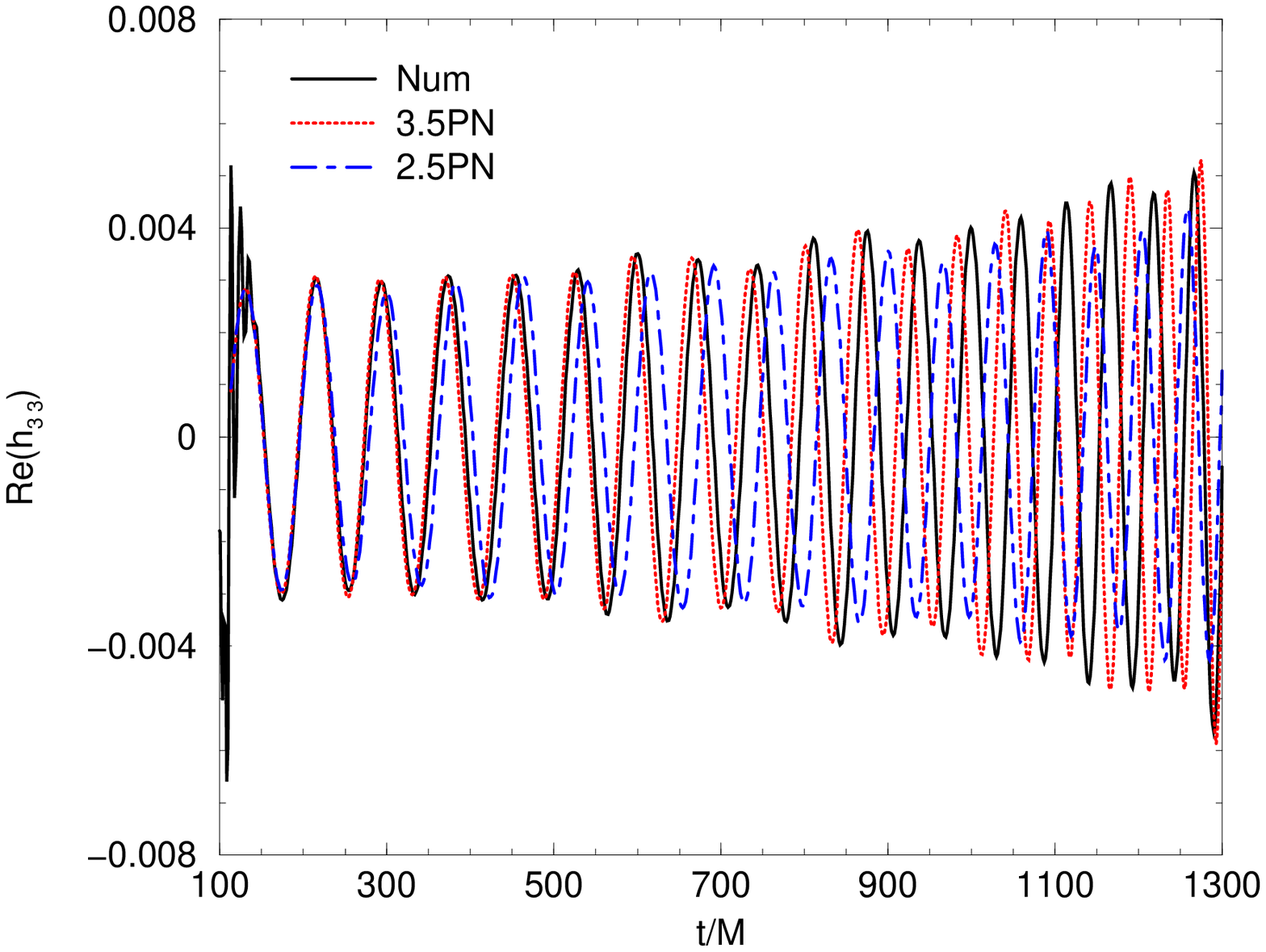}
\caption{
The real part of the $(\ell=3, m=3)$ mode of $h$ for the G2.5
configuration from the numerical, truncated 2.5PN, and 3.5PN simulations. 
Note the relatively high-frequency oscillations in the amplitude
(roughly corresponding to the orbital period).
}
\label{fig:G2.5_re_h33}
\end{figure}

\begin{figure}
\includegraphics[width=3.5in]{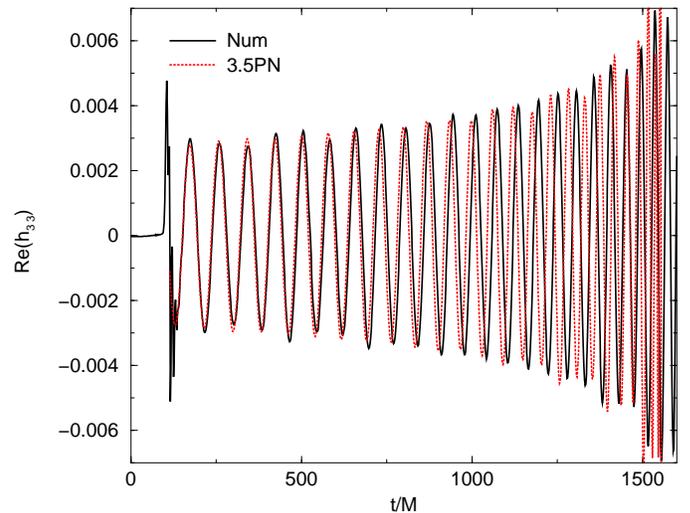}
\caption{The real part of the $(\ell=3, m=3)$ mode of $h$ for the G3.5
configuration from the numerical and 3.5PN simulations. Note the
relatively high-frequency oscillations in the amplitude
(roughly corresponding to the orbital period).}
\label{fig:G3.5_re_h33}
\end{figure}

We produced both 3.5PN and 2.5 truncated PN waveforms for the G2.5
configuration and 3.5PN waveforms for the G3.5 configuration. In
Figs.~\ref{fig:G2.5_re_h22}~and~\ref{fig:G3.5_re_h22}, we show the
real part of the $(\ell=2, m=2)$ mode of the strain $h$ for G2.5 and G3.5
respectively. Note the reasonable
agreement of the numerical and 3.5PN waveforms for $700M$ in both
configurations.  The differences between the PN and numerical
waveforms are larger than the numerical waveform errors at this time.
Also note that the 3.5PN waveform shows evidence of an early merger
and has a higher frequency than
the numerical waveform, while 2.5PN waveform shows the opposite behavior.  In
Figs.~\ref{fig:G2.5_re_h21}~and~\ref{fig:G3.5_re_h21}, we show the
real part of the $(\ell=2, m=1)$ mode of $h$ for G2.5 and G3.5
respectively. Again, the agreement is
fairly good at earlier times and 3.5PN is more accurate than 2.5PN.
Also, note the interesting oscillatory behavior of the amplitude of
the real part of the $(\ell=2,m=1)$ mode for
both configurations.
Here the amplitude (of the real part) oscillates at
about the precessional frequency (see Fig.~\ref{fig:G3.5_Precess}).
For the $(\ell=3,m=3)$ mode, we obtained 
results similar to the
$(\ell=2,m=2)$ mode,  as seen in
Figs.~\ref{fig:G2.5_re_h33}~and~\ref{fig:G3.5_re_h33}. However, for
this mode, oscillations in the amplitude are more pronounced.




\subsubsection{Amplitudes}
\label{sec:pn_amp}

We are concerned with exploring two different effects, eccentricity
and precession. Long-term precessional effects, which modulate the
amplitude of the waveform over many cycles, are more readily apparent
in $h$ because differentiating $h$ twice (to obtain $\psi_4$) suppresses
low-frequency oscillations in comparison to higher frequencies. As the
binary inspirals, the frequency of the oscillations increases with the
orbital frequency. Thus there is a large ramp-up in the amplitude of $\psi_4$
near merger. This can mask other effects as we observe below. On the
other hand, the transformation from $\psi_4$ to $h$ can induce both
high-frequency and low-frequency distortions in $h$ (i.e.\ numerical
errors due to the windowing procedure in the Fourier transform). Thus
it is advantageous to compare both $\psi_4$ and $h$ between the PN and
numerical simulations.

\begin{figure}[ht]
\includegraphics[width=3.5in]{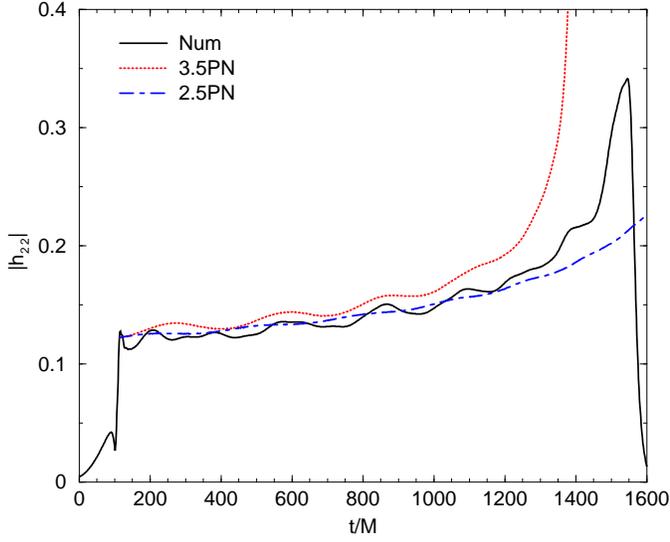}
\caption{The amplitude of the $(\ell=2, m=2)$ mode of $h$ for the
G2.5 configuration from the numerical, truncated 2.5PN, and 3.5PN simulations. 
The oscillations in the amplitude are much more pronounced in the
numerical and 3.5PN simulations, indicating that these oscillations
are likely due to eccentricity.
}
\label{fig:G2.5_AMP_h22}
\end{figure}

\begin{figure}
\includegraphics[width=3.5in]{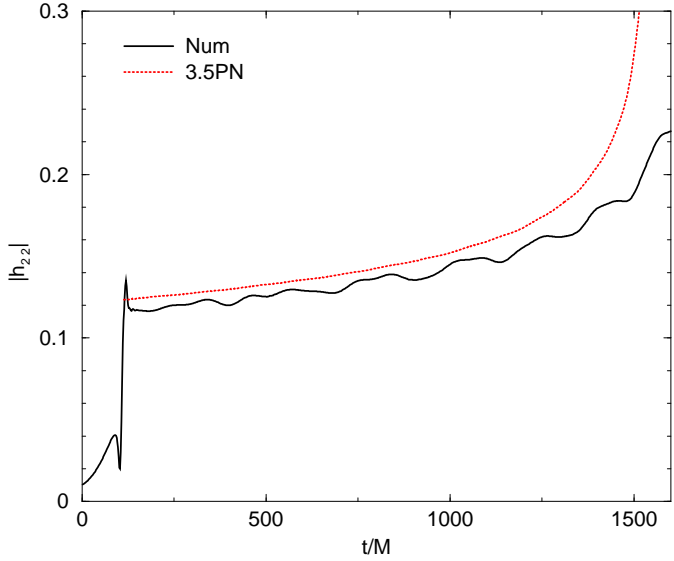}
\caption{The amplitude of the $(\ell=2, m=2)$ mode of $h$ for the
G3.5 configuration from the numerical and 3.5PN simulations.
The amplitude oscillations in the numerical waveform are much larger
than those in the 3.5PN waveform, indicating that they are likely due
to eccentricity
}
\label{fig:G3.5_AMP_h22}
\end{figure}

\begin{figure}
\includegraphics[width=3.5in]{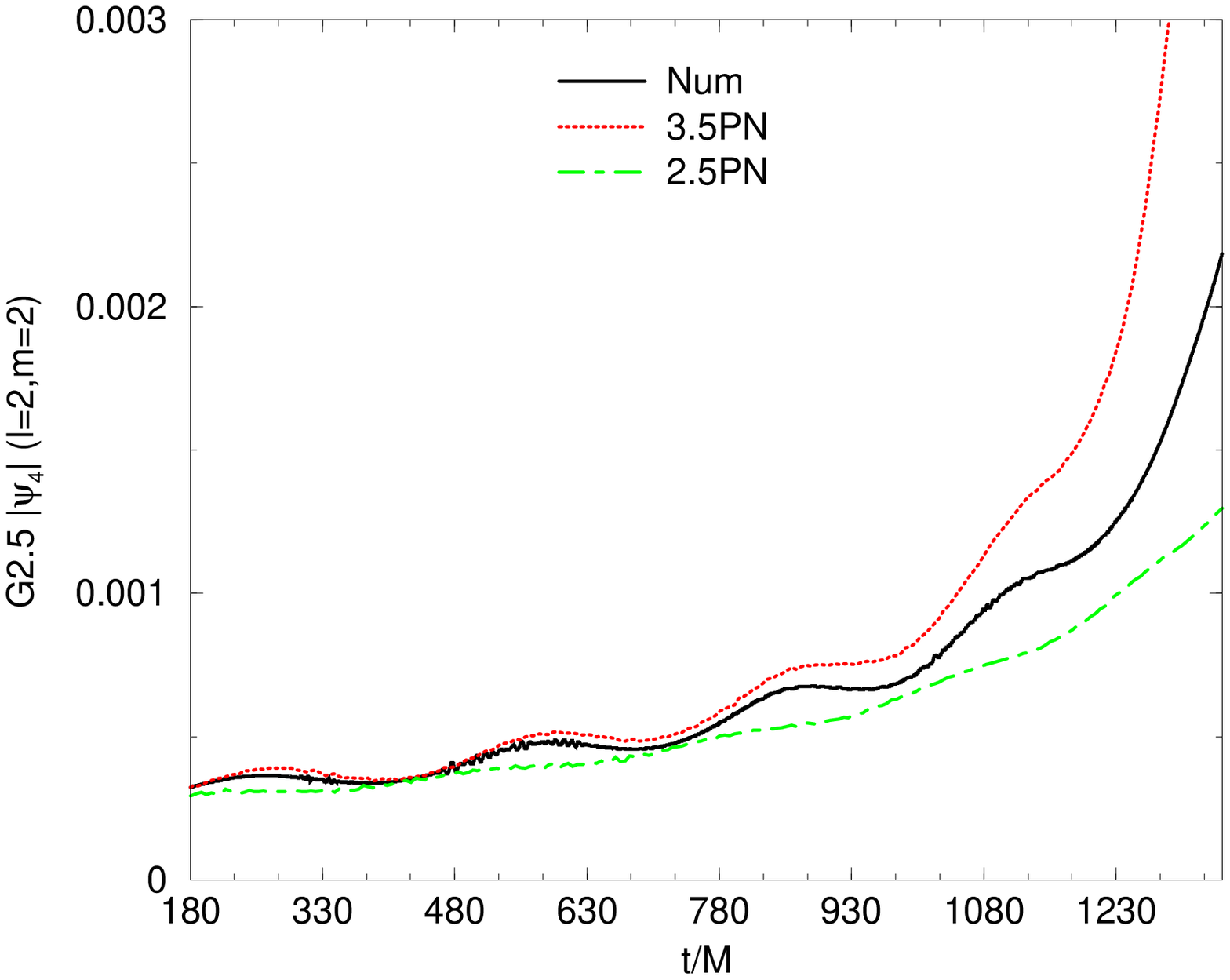}
\caption{The amplitude of the $(\ell=2, m=2)$ mode of $\psi_4$ for the
G2.5 configuration from the numerical, truncated 2.5PN, and 3.5PN simulations. 
Note the very good agreement between the 3.5PN and numerical
waveforms.
}
\label{fig:G2.5_amp_p4_2_2}
\end{figure}

\begin{figure}
\includegraphics[width=3.5in]{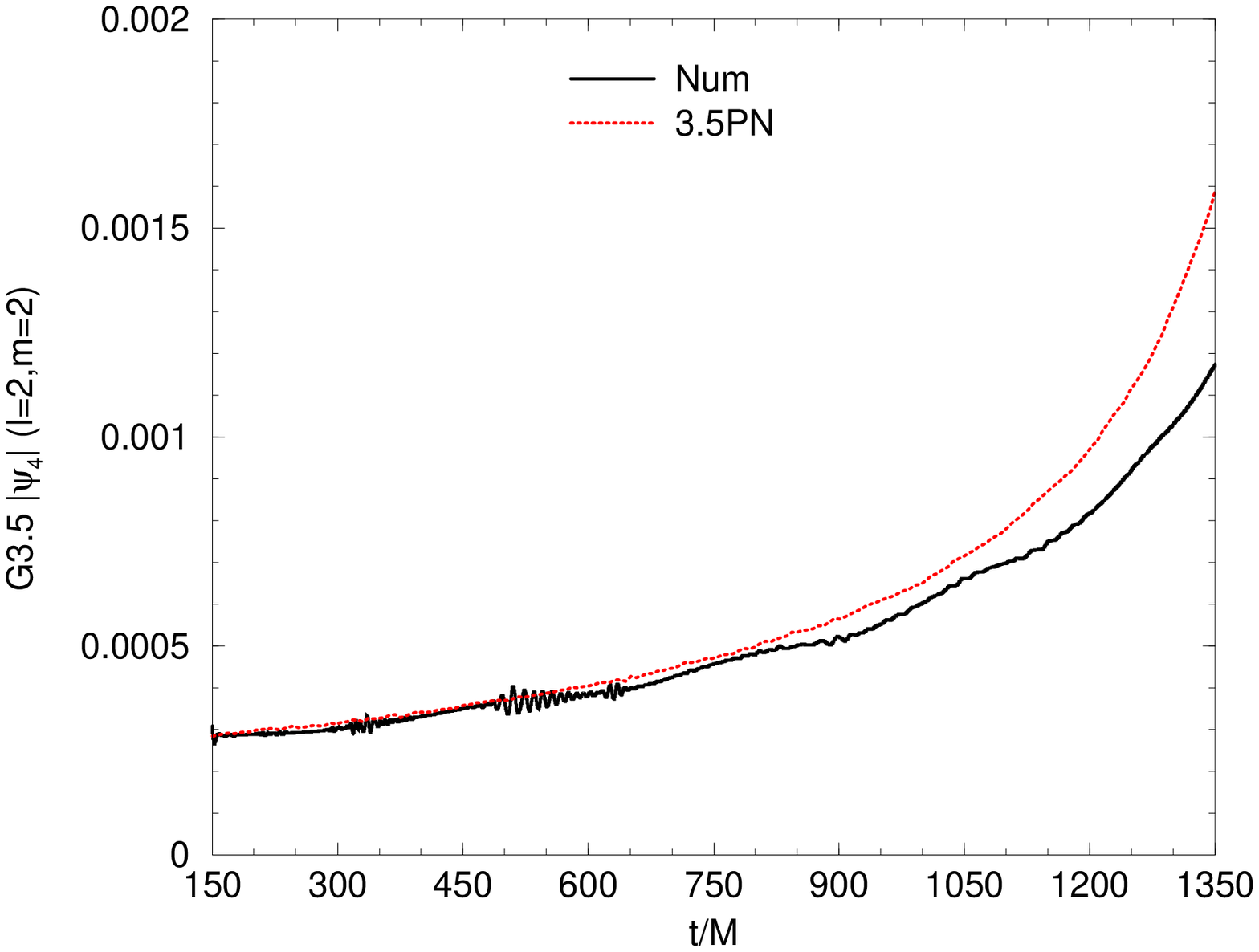}
\caption{The amplitude of the $(\ell=2, m=2)$ mode of $\psi_4$ for the
G3.5 configuration from the numerical and 3.5PN simulations. 
}
\label{fig:G3.5_amp_p4_2_2}
\end{figure}

In order to analyze the behavior of the $(\ell,m)$ modes of the
waveform, we decompose the modes into amplitudes and phases. In
Fig.~\ref{fig:G2.5_AMP_h22} we show the amplitude of the
$(\ell=2,m=2)$ mode of $h$ for the G2.5 configuration. Here the 2.5PN
waveforms appear to capture the overall amplitude behavior to better
accuracy, while the 3.5PN waveforms capture the oscillations in the
amplitude. These oscillations occur at roughly the orbital
frequency and are due mainly to eccentricity and, to a lesser
extent, precession. As discussed above, precession can induce an
oscillation in the
$(\ell=2,m=2)$ mode at the orbital frequency by mixing the $(\ell =2,
m=2)$ and $(\ell=2,m=\pm1)$ modes (and since the
$m$ modes have frequency $\sim|m|\omega_{\rm orbit}$, where
$\omega_{\rm orbit}$ is the orbital
frequency, the resulting modes will show a beating effect at the
orbital frequency).
 A similar plot for the G3.5 configuration,
Fig.~\ref{fig:G3.5_AMP_h22}, shows that 3.5PN
predicts very small amplitude oscillations, which seem to confirm that
the oscillations seen in G2.5 are mainly due to eccentricity.
Note that in Fig.~\ref{fig:G3.5_AMP_h22} the amplitude of the
numerical $(\ell=2,m=2)$ mode oscillates at about the orbital
frequency with a significantly larger amplitude than the 3.5PN
prediction; indicating that these oscillations are due to eccentricity
(which is consistent with the relatively large oscillations in the
numerical orbital radius). 
Since the  transformation from $\psi_4$ to $h$ can induce artifacts
into the waveforms, it is also important to
compare the PN predictions for $\psi_4$ with the numerical waveforms.
In Figs.~\ref{fig:G2.5_amp_p4_2_2}~and~\ref{fig:G3.5_amp_p4_2_2} we
show the amplitude of the $(\ell=2,m=2)$ of $\psi_4$ for the G2.5 and
G3.5 configurations respectively. Note that, for $\psi_4$, 3.5PN gives
a clearly better fit to the G2.5 waveform than truncated 2.5PN.
Note also that the agreement between the 3.5PN and numerical $\psi_4$
appears to be significantly better than the agreement in $h$. Thus it
appears that the windowing procedure has induced a very-low frequency mode
into $h$ that yielded a net change in the amplitude of the waveform.

\begin{figure}[ht]
\includegraphics[width=3.5in]{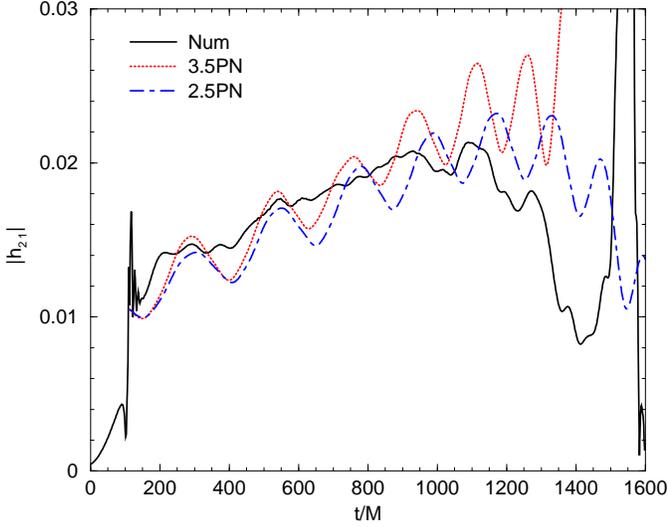}
\caption{The amplitude of the $(\ell=2, m=1)$ mode of $h$ for the
G2.5 configuration from the numerical, truncated 2.5PN, and 3.5PN
simulations. The secular oscillation in the numerical amplitude
occurs at roughly the precessional frequency. Here the
shorter-timescale oscillations apparent in the PN waveforms are much
smaller in the numerical waveform.
}
\label{fig:G2.5_AMP_h21}
\end{figure}

\begin{figure}
\includegraphics[width=3.5in]{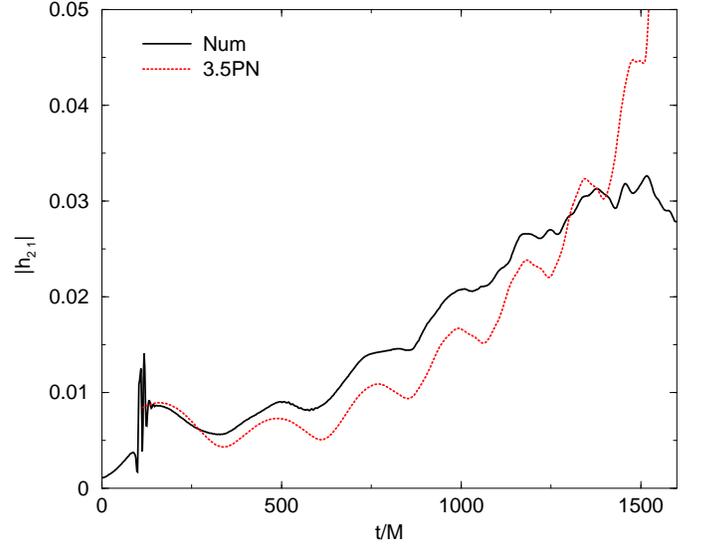}
\caption{The amplitude of the $(\ell=2, m=1)$ mode of $h$ for the
G3.5 configuration from the numerical and 3.5PN simulations.
The secular oscillation in the numerical amplitude
occurs at roughly the precessional frequency. Here the
shorter-timescale oscillations (corresponding roughly to the orbital
period) are present in both waveforms with very similar amplitudes.
}
\label{fig:G3.5_AMP_h21}
\end{figure}

\begin{figure}
\includegraphics[width=3.5in]{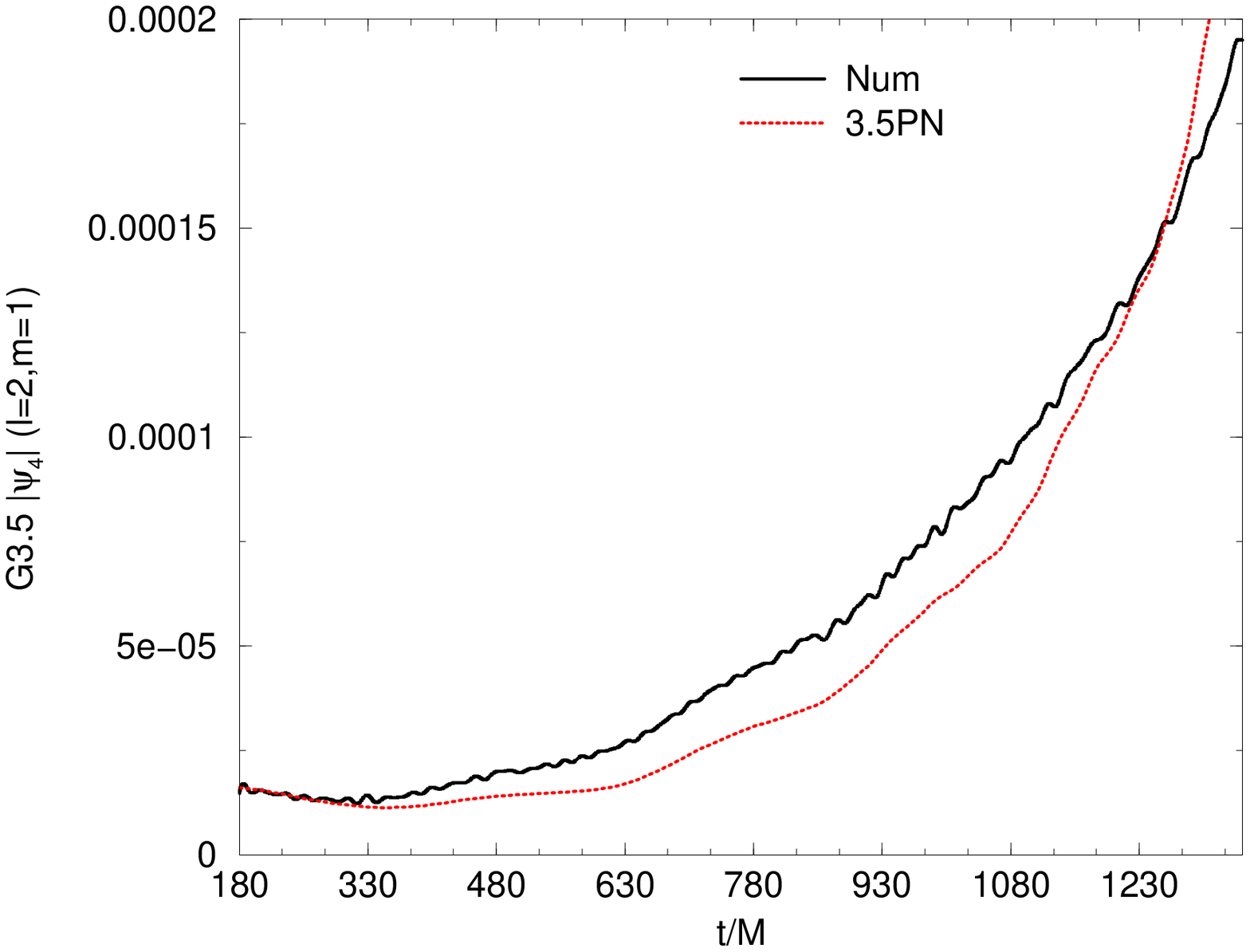}
\caption{The amplitude of the $(\ell=2, m=1)$ mode of $\psi_4$ for the
G3.5 configuration from the numerical and 3.5PN simulations. 
}
\label{fig:G3.5_amp_p4_2_1}
\end{figure}

The effects of precession become apparent in the sub-leading modes $h$
(and to a lesser extent, in the sub-leading modes of $\psi_4$).
However, numerical errors in the lower amplitude modes are also more
pronounced. In Fig.~\ref{fig:G2.5_AMP_h21}~and~\ref{fig:G3.5_AMP_h21}
we show the amplitudes of the $(\ell=2,m=1)$ mode of $h$ for the G2.5
and G3.5 configurations, respectively. Here both 2.5PN and 3.5PN
capture the secular behavior in the amplitude nicely. Unlike for the
$(\ell=2,m=2)$ mode, here the PN amplitudes oscillate much more
strongly than the numerical amplitudes for the G2.5 configuration,
while 3.5PN seems to capture both the short (orbital frequency)
timescale oscillations and the longer (precessional) frequency
oscillation (until $t\sim 1000M$) for the G3.5 configuration.  The
damping of the numerical oscillations for the G2.5 configuration are
likely a consequence of the windowing procedure (which acts as
a high-frequency and low-frequency filter), as a similar  
damping is not apparent in
 $\psi_4$ (See
Figs.~\ref{fig:G3.5_AMP_h21}~and~\ref{fig:G3.5_amp_p4_2_1}).  
Although the G3.5 configuration has very low eccentricity (according
to 3.5PN), the effects of eccentricity can increase as the binary
separation falls below $15M$ (See Fig.~\ref{fig:pn_ecc_gen}).
This effect appears to be related to precession because the
eccentricity of non-precessing binaries (See Fig.~\ref{fig:pn_ecc_up_up})
decreases uniformly with binary separation.
In addition, mode-mixing effects may also be partially responsible for
these oscillations in the amplitude of the $(\ell=2,m=1)$ mode at the
orbital frequency.
 The secular oscillation in
the amplitude of the $(\ell=2,m=1)$ mode matches the precessional
frequency (See Figs.~\ref{fig:G3.5_rotate}~and~\ref{fig:G3.5_AMP_h21}),
 and is thus likely a direct consequence of precession (the
amplitude of the $(\ell=2,m=1)$ mode contains significant
contributions from the spins, see Eq.~(3) in~\cite{Berti:2007nw}).

\begin{figure}
\includegraphics[width=3.5in]{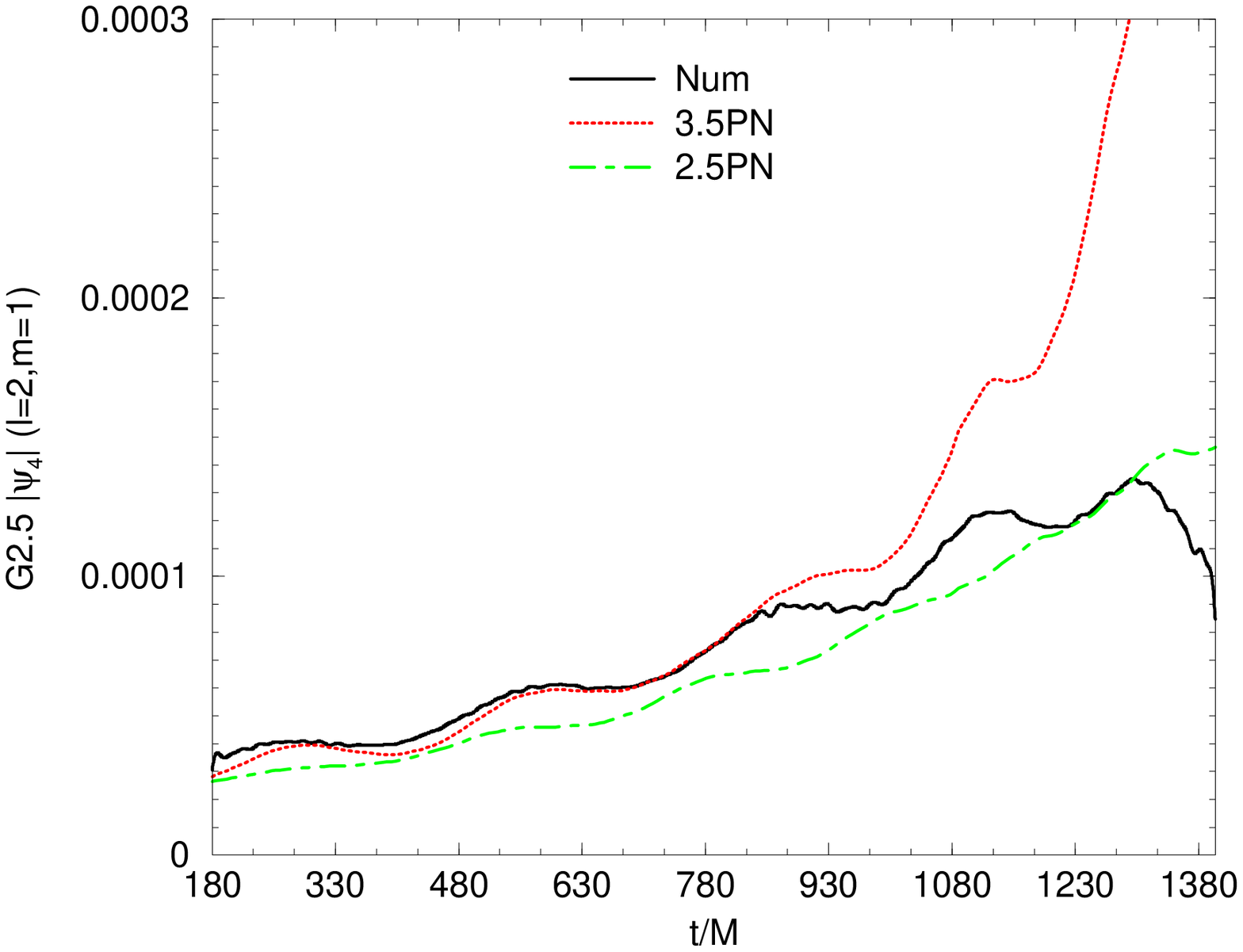}
\caption{The amplitude of the $(\ell=2, m=1)$ mode of $\psi_4$ for the
G2.5 configuration from the numerical, truncated 2.5PN, and 3.5PN simulations. 
Note the very good agreement between the 3.5PN and numerical
waveforms.
}
\label{fig:G2.5_amp_p4_2_1}
\end{figure}

The $(\ell=2,m=1)$ mode of $\psi_4$, as seen in
Figs.~\ref{fig:G2.5_amp_p4_2_1}~and~\ref{fig:G3.5_amp_p4_2_1} again
shows that the 3.5PN waveforms are clearly more accurate than the
truncated 2.5PN waveforms. The agreement of the 3.5PN waveforms for
the G2.5 configuration is remarkable. 
Note that the long-timescale oscillation
seen in the $(\ell=2,m=1)$ mode of $h$,
which is likely due to precession, is not apparent in $\psi_4$ of the
G3.5 configuration. However, as this effect is smaller in G3.5
(as seen by comparing
Figs.~\ref{fig:G2.5_amp_p4_2_1}~and~\ref{fig:G3.5_amp_p4_2_1}),
it may be hidden
in $\psi_4$ by the ramp-up in amplitude of $\psi_4$ near merger.

\begin{figure}[ht]
\includegraphics[width=3.5in]{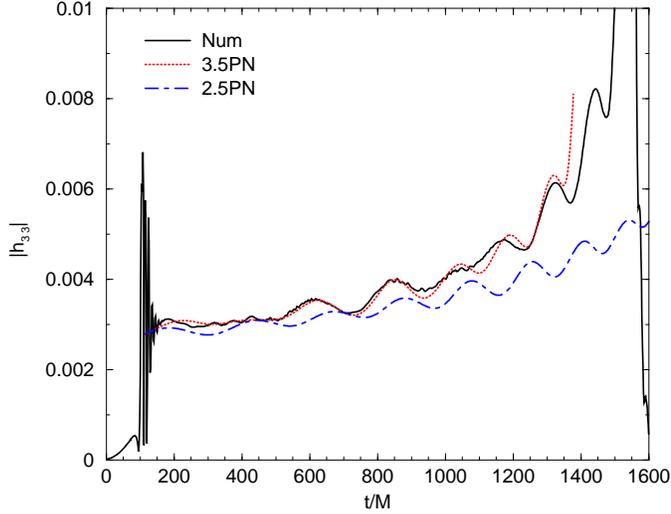}
\caption{The amplitude of the $(\ell=3, m=3)$ mode of $h$ for the
G2.5 configuration from the numerical, truncated 2.5PN, and 3.5PN simulations. 
Note the very good agreement between the 3.5PN and numerical
waveforms. Also note that the short-timescale oscillations (orbital
period) grow with time at later times, indicating that, at least at
later times, they are due mainly to precession.
}
\label{fig:G2.5_AMP_h33}
\end{figure}

\begin{figure}
\includegraphics[width=3.5in]{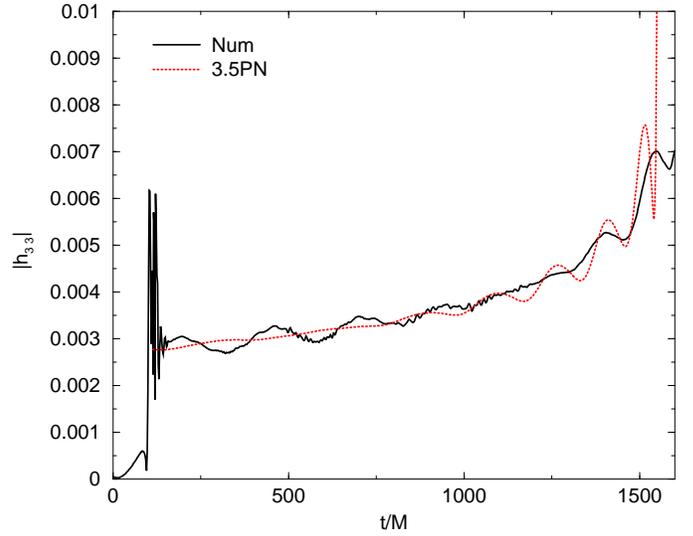}
\caption{The amplitude of the $(\ell=3, m=3)$ mode of $h$ for the
G3.5 configuration from the numerical and 3.5PN simulations. Note that
the short-timescale oscillation at later times grow with time,
indicating that these later-time oscillations are due to precession.
The early-time oscillations in the numerical waveform (at the same
frequency) are likely due to eccentricity.
}
\label{fig:G3.5_AMP_h33}
\end{figure}

\begin{figure}
\includegraphics[width=3.5in]{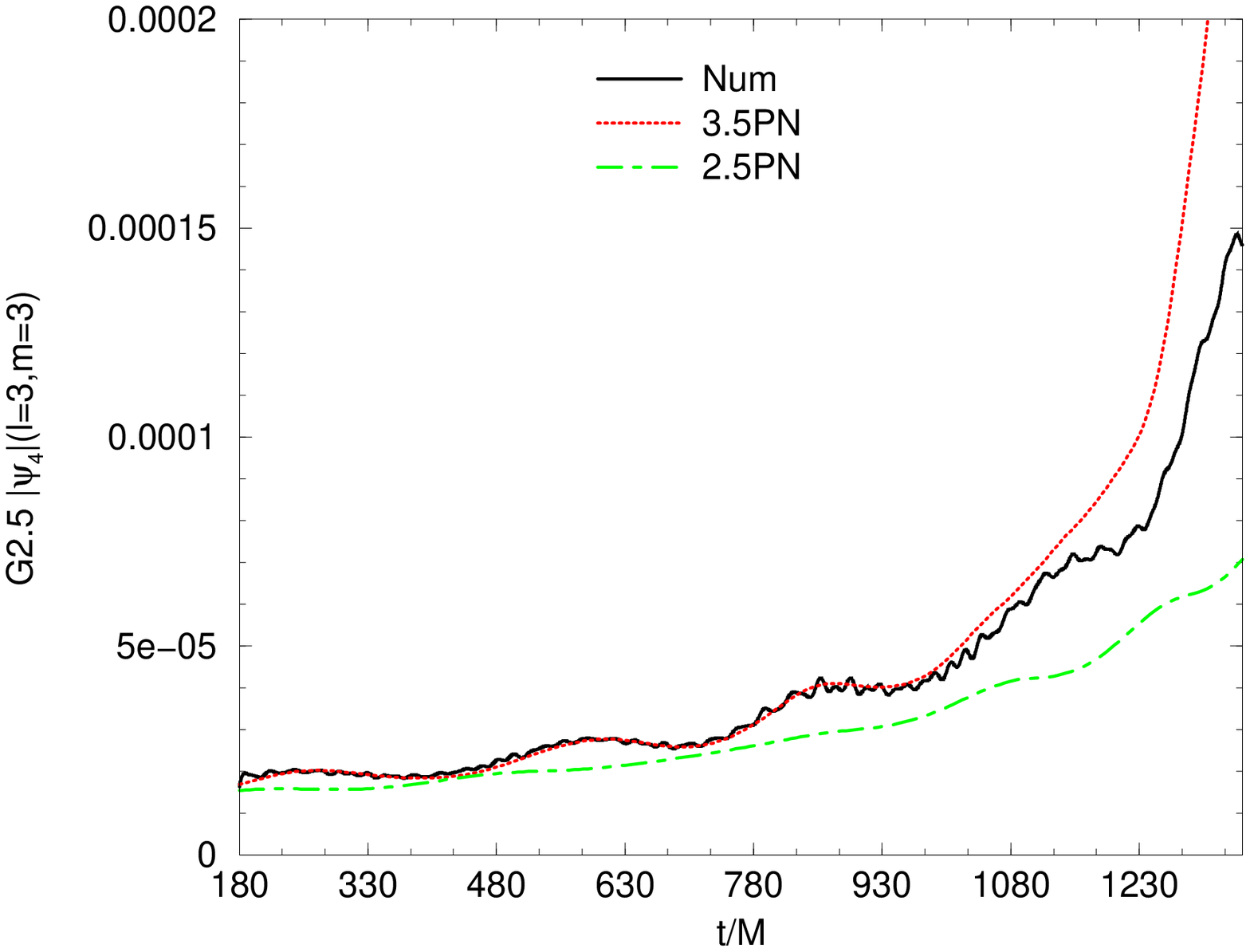}
\caption{The amplitude of the $(\ell=3, m=3)$ mode of $\psi_4$ for the
G2.5 configuration from the numerical, truncated 2.5PN, and 3.5PN simulations. 
Note the very good agreement between the 3.5PN and numerical
waveforms.
}
\label{fig:G2.5_amp_p4_3_3}
\end{figure}

\begin{figure}
\includegraphics[width=3.5in]{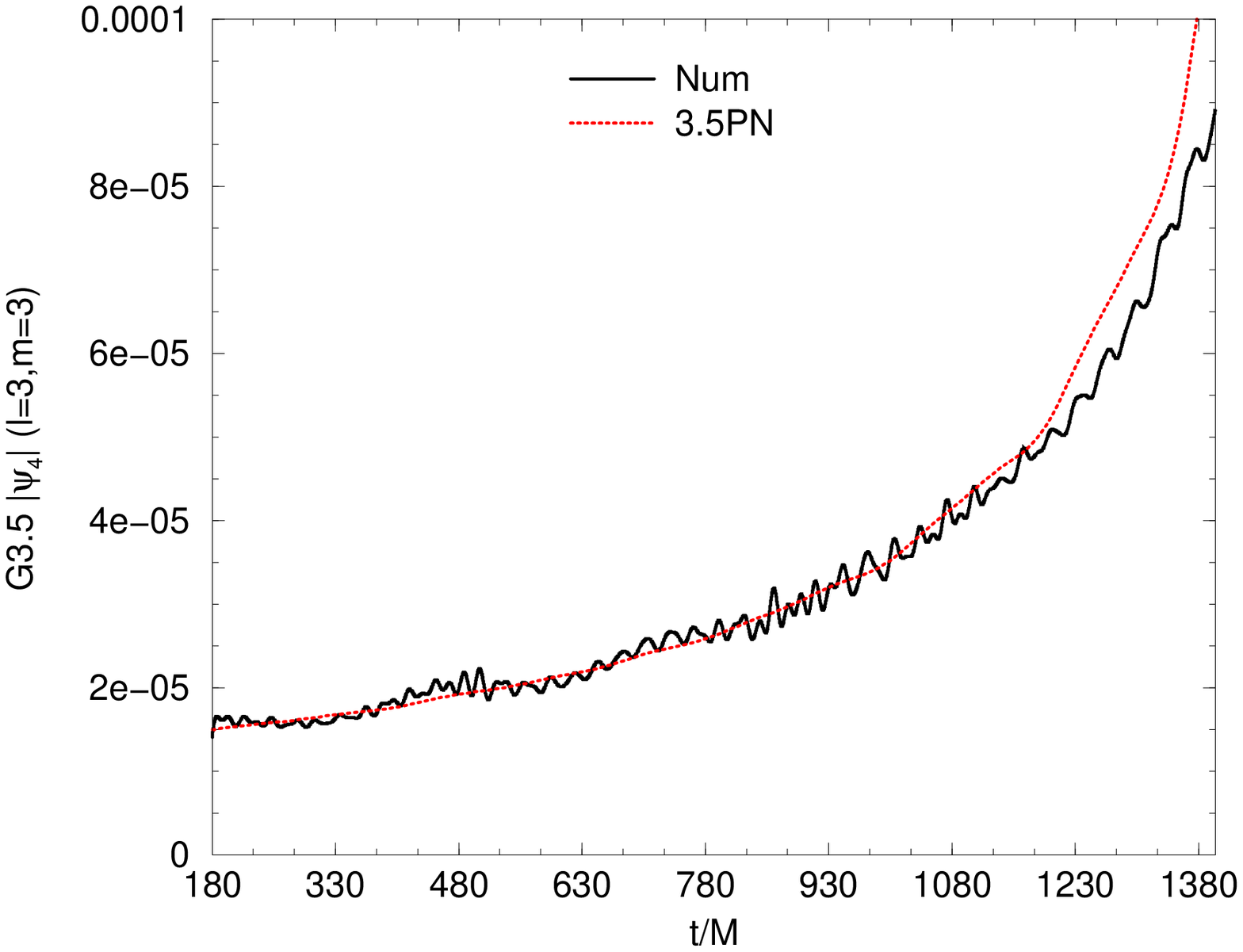}
\caption{The amplitude of the $(\ell=3, m=3)$ mode of $\psi_4$ for the
G3.5 configuration from the numerical and 3.5PN simulations. 
}
\label{fig:G3.5_amp_p4_3_3}
\end{figure}

Finally, in Fig.~\ref{fig:G2.5_AMP_h33}~and~\ref{fig:G3.5_AMP_h33} we
show the amplitudes of the $(\ell=3,m=3)$ mode of $h$ for the G2.5 and
G3.5 configurations, respectively.  An interesting feature of these
modes is that the late-time amplitude oscillations, which are roughly
at the orbital frequency, increase with time, indicating that they are
due to the precession-induced late-time eccentricity apparent
in Fig.~\ref{fig:pn_ecc_gen}
For the G2.5
configuration, 3.5PN produces a remarkably good fit, capturing all
oscillations in the amplitude until $t\sim1400M$. On the other hand,
3.5PN does not capture the early-time oscillations in the G3.5
configuration. A possible explanation for this result is that, as seen
in Figs.~\ref{fig:G2.5_PN_NUM_R}~and~\ref{fig:G3.5_PN_NUM_R}, both
3.5PN and the numerical simulation show similar eccentricities for the
G2.5 configuration, but 3.5PN shows much lower eccentricity for
the G3.5 configuration. This eccentricity leads to the early-time
oscillation in the amplitude of the $(\ell=3,m=3)$ mode that are not
captured by 3.5PN. However, as the binary evolves, the effects of 
precession-induced eccentricity in the PN EOM increase and eventually
dominate. This
causes the amplitude of the oscillations in the 3.5PN waveform to
increase and eventually become larger than the numerical amplitude
oscillations.
In Figs.~\ref{fig:G2.5_amp_p4_3_3}~and~\ref{fig:G3.5_amp_p4_3_3} we
show the amplitude of the $(\ell=3,m=3)$ mode of $\psi_4$ for the G2.5
and G3.5 configurations. Here too 3.5PN gives a remarkably good
estimation for the amplitude of the mode. Note that the
orbital-frequency oscillations seen in Fig.~\ref{fig:G3.5_AMP_h33}
are not readily apparent in
Fig.~\ref{fig:G3.5_amp_p4_3_3} (even in the PN waveforms).
This shows one advantage of analyzing 
$h$ over $\psi_4$; eccentricity and precessional effects are more
apparent in $h$.

From the amplitudes of each mode, we see that precession and
eccentricity impart signatures on the modes of the waveform at the
orbital frequency. However, the long-time oscillations in the
amplitudes, here apparent only in the $(\ell=2,m=\pm1)$ modes, seem to
be due purely to precession, and occur at the precessional frequency.




\subsubsection{Phases}

\begin{figure}[ht]
\includegraphics[width=3.5in]{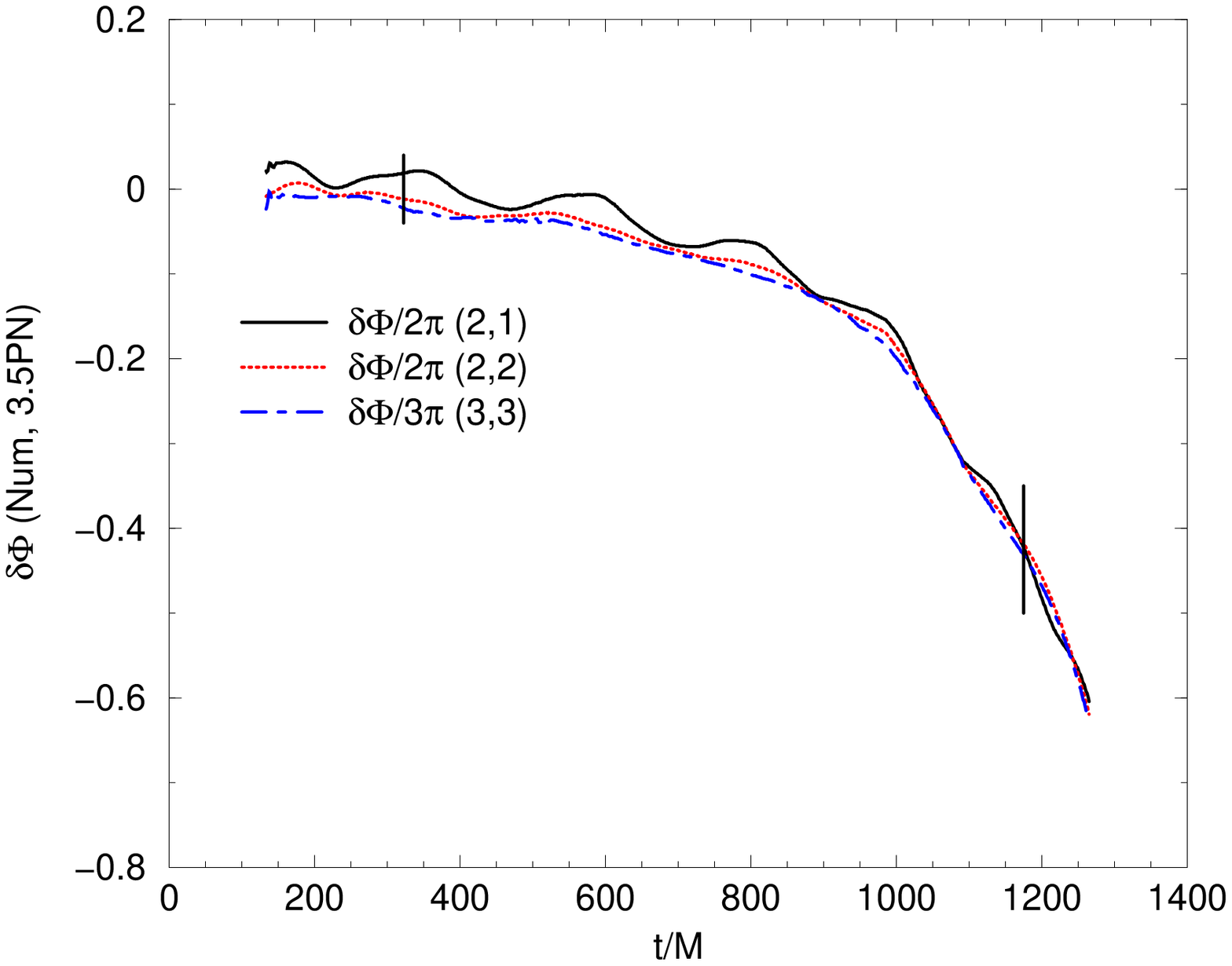}
\caption{The phase differences in $h$ between the numerical and
3.5 PN
simulations for the G2.5 configuration in the $(\ell=2,m=1)$,
$(\ell=2,m=2)$, and $(\ell=3,m=3)$ modes. We multiplied the phase
differences in the modes by a factor of $1/(\ell \pi)$. We divide by
$\ell \pi$, rather than $m \pi$, because the $(\ell=2,m=1)$ mode is
dominated by mode-mixing from the $(\ell=2,m=\pm2)$ modes (see text
for more details). The vertical lines shows the times when the
$(\ell=2,m=2)$ frequency is $M\omega = 0.05$ ($t\sim323M$) and $M\omega = 0.075$
($t\sim1075M$).
}
\label{fig:G2.5_PHASE_ERR}
\end{figure}

\begin{figure}
\includegraphics[width=3.5in]{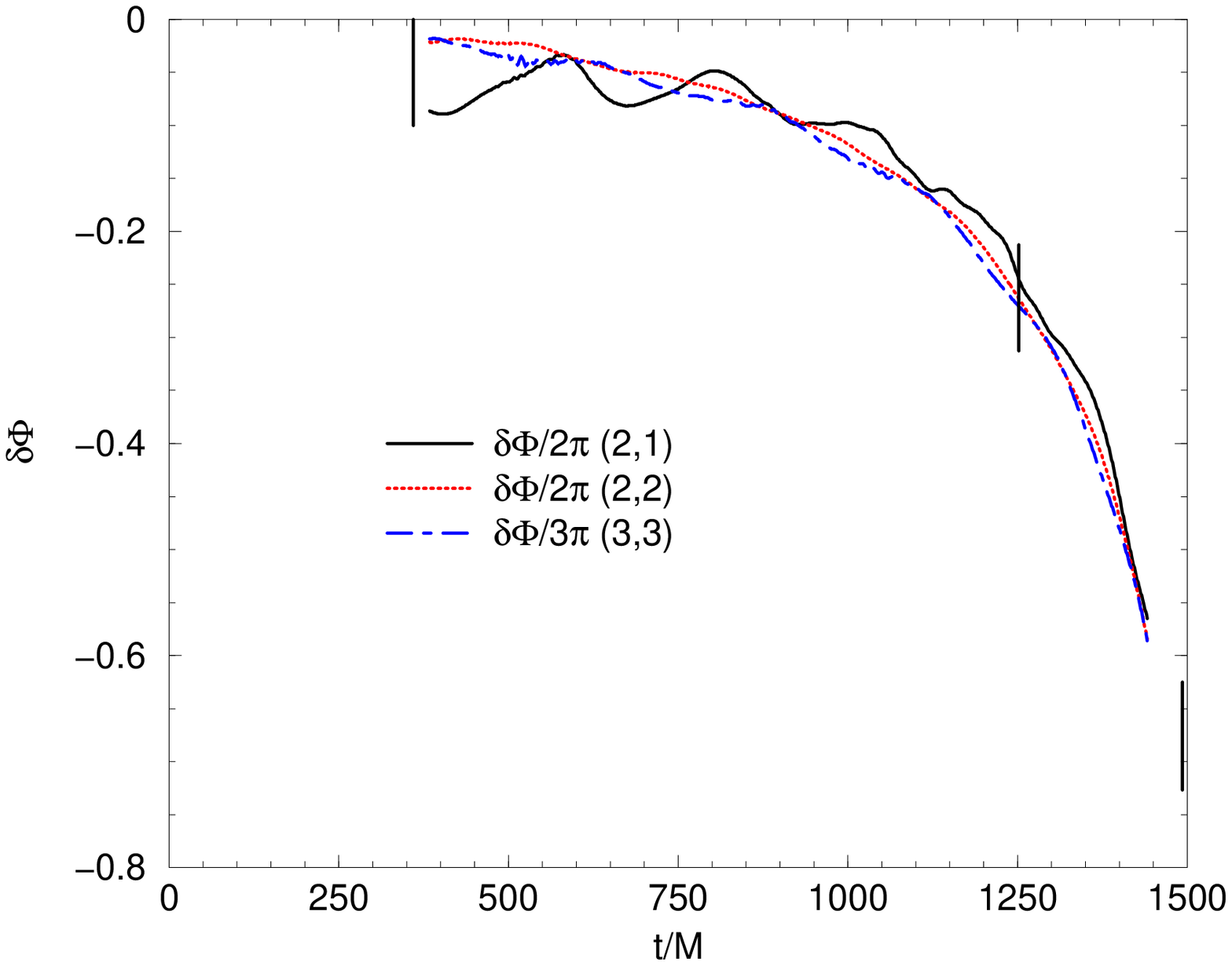}
\caption{The phase differences in $h$ between the numerical and 
3.5 PN
simulations for the G3.5 configuration in the $(\ell=2,m=1)$,
$(\ell=2,m=2)$, and $(\ell=3,m=3)$ modes. 
 We multiplied the phase
differences in the modes by a factor of $1/(\ell \pi)$.
Note that the normalized phase differences are qualitatively
independent of the mode and arise from the orbital
phase error in the PN approximation.
We divide by
$\ell \pi$, rather than $m \pi$, because the $(\ell=2,m=1)$ mode is
dominated by mode-mixing from the $(\ell=2,m=\pm2)$ modes (see text
for more details). The vertical lines show the times when the
$(\ell=2,m=2)$ frequency is $M\omega = 0.05$ ($t\sim360M$),
$M\omega = 0.075$ ($t\sim1252M$), and $M\omega = 0.1$ ($t\sim1493M$).
}
\label{fig:G3.5_PHASE_ERR}
\end{figure}

\begin{figure}[ht]
\includegraphics[width=3.5in]{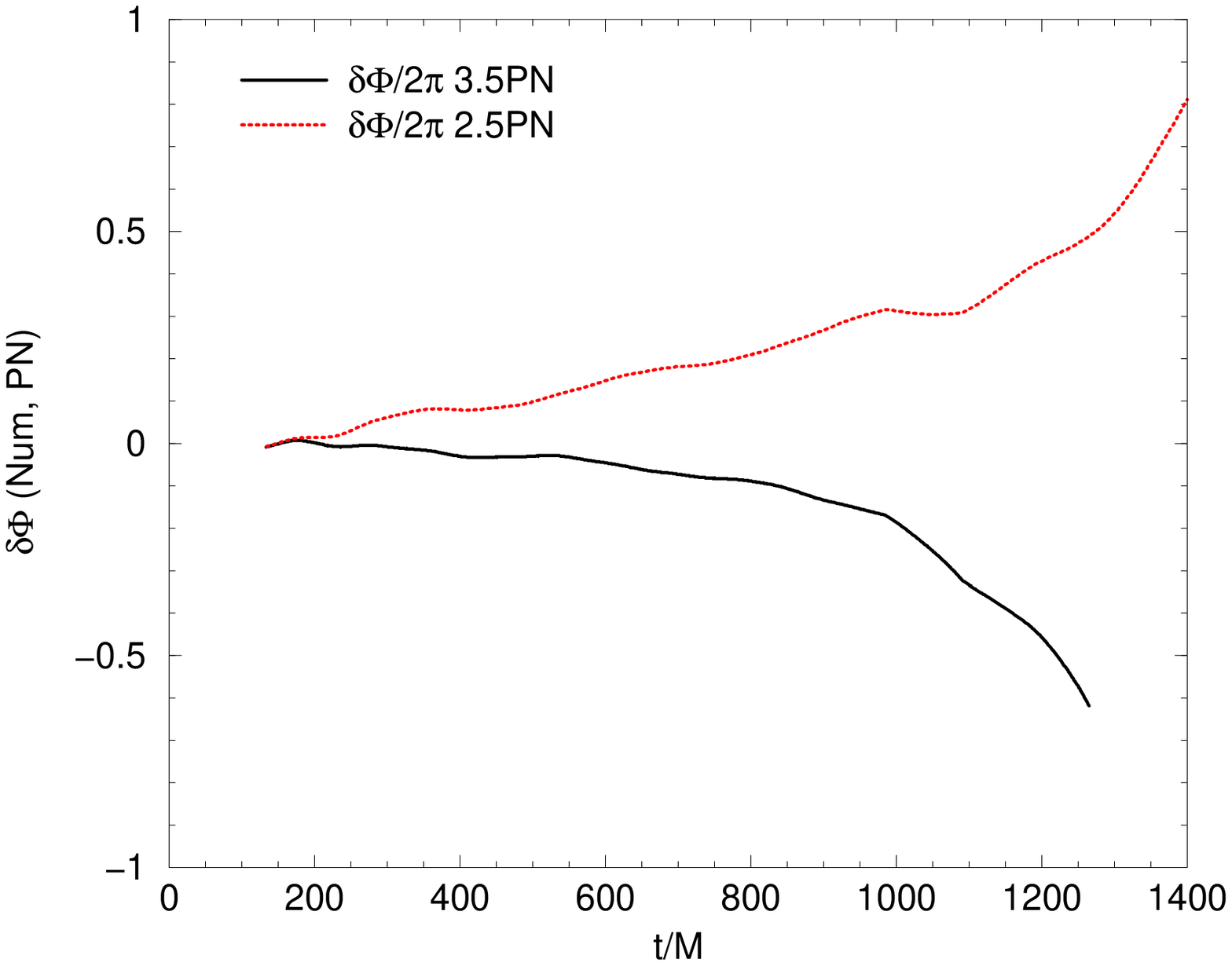}
\caption{The phase difference in the $(\ell=2,m-2)$ mode of $h$
between the NR and PN waveforms
for the G2.5 configuration. 
The vertical axis denotes the number of orbital rotations derived from
GW cycle. Note that the normalized phase differences are qualitatively
independent of the mode and arise from the orbital
phase error in the PN approximation.
}
\label{fig:G2.5_PHASE_CMP}
\end{figure}

In Figs.~\ref{fig:G2.5_PHASE_ERR}~and~\ref{fig:G3.5_PHASE_ERR} we show the
phase differences between the 3.5PN and numerical waveforms for
the $(\ell=2,m=1)$, $(\ell=2,m=2)$, and $(\ell=3,m=3)$ modes. In all
cases we normalized the phase differences by dividing by $\ell \pi$.
Note that we renormalize by $\ell \pi$, rather than $m \pi$.
If the orbital plane were to lie along the $xy$ plane, or
equivalently, we chose spherical coordinates such that the $\theta=0$
corresponds to direction of normal to the orbital plane, then we would
expect the $(\ell, m)$ modes to have frequency $\omega\approx m\ \omega_{\rm
orbit}$, and an error in the orbital phase of $\delta\Phi_{\rm orbit}$
would lead to an error in the phase of the $(\ell, m)$ modes of
$m\ \delta\Phi_{\rm orbit}$. However, in that case the $(\ell=2, m=1)$
mode would be very small.  Consequently, in our non-aligned spin
basis, the $(\ell=2,m=1)$ mode is actually dominated by contributions
from the $(\ell=2,m=\pm2)$ modes (of the aligned spin-basis). Thus, in
our configurations, the $(\ell=2,m=1)$ mode has frequency $2
\omega_{\rm orbit}$ and the error in the phase scales like $2 \delta
\Phi_{\rm orbit}$.
Note that the renormalized phase differences are qualitatively
independent of the mode.
We therefore focus on the $(\ell=2,m=2)$ mode. In
Fig.~\ref{fig:G2.5_PHASE_CMP} we show the phase difference between the
numerical, 2.5PN, and 3.5PN $(\ell=2,m=2)$ mode of $h$ for the
G2.5 configuration. From the plot we see that the phase difference
improves with the higher PN order and changes sign.
It thus appears that still higher-order PN corrections may make the
waveform phases agree. 
As seen in Figs.~\ref{fig:G2.5_re_h21},~\ref{fig:G2.5_re_h22},
and~\ref{fig:G2.5_re_h33}, the truncated 2.5PN phase evolution 
is slower than that of the NR and 3.5PN, and thus its phase lags
behind the other two. The 3.5 PN evolution merges too quickly (but is
still closer to the numerical evolution) and thus its phase leads
the numerical one.

\subsubsection{Matching}

In order to quantitatively compare the modes of the truncated 2.5PN
and 3.5PN waveforms with the numerical waveforms we define the
overlap, or matching criterion, for the real and imaginary parts of
each mode as
\begin{eqnarray}
  \label{eq:match}
  M_{\ell m}^\Re &=& \frac{<R^{\rm Num}_{\ell m},
    R^{\rm PN}_{\ell m}>}
     {\sqrt{<R^{\rm Num}_{\ell m},R^{\rm Num}_{\ell m}>
     <R^{\rm PN}_{\ell m},R^{\rm PN}_{\ell m}>}},\\
  M_{\ell m}^\Im &=& \frac{<I^{\rm Num}_{\ell m},
    I^{\rm PN}_{\ell m}>}{\sqrt{<I^{\rm Num}_{\ell m},
    I^{\rm Num}_{\ell m}><I^{\rm PN}_{\ell m},I^{\rm PN}_{\ell m}>}},
\end{eqnarray}
where
$R_{\ell m} = Re(h_{\ell m})$,
$I_{\ell m} = Im(h_{\ell m})$, and
\begin{equation}
<f,g> = \int_{t_1}^{t_2} f(t) g(t) dt.
\end{equation}
Hence, $M_{\ell m}^\Re = M_{\ell m}^\Im = 1$ indicates that the given
PN and numerical mode agree.
To compare PN and numerical waveforms, we need to determine
the time translation $\delta t$ between the numerical time and the
corresponding point on the PN trajectory. That is to say, the time it
takes for the signal to reach the extraction sphere (here $r=100M$).
We determine this time translation by finding the time translation
near $\delta t=100M$ that maximizes the agreement of the early time
waveforms in the $(\ell=2,m=\pm2)$, $(\ell=2,m=\pm1)$, and
$(\ell=3,m=\pm3)$ simultaneously. We find $\delta t \sim 112$, in
good agreement with the expectation for our observer at $r=100M$. We
also determine an alternate time translation, one full wavelength in
the $(\ell=2,m=2)$ mode longer, that increases the matching of the
$(\ell=2,m=2)$ mode over longer integration periods. On the other
hand, this new time
translation, $\delta t = 233$, causes the $(\ell=3)$ modes to be out
of phase, leading to negative overlaps. Thus by looking at the
$(\ell=2)$ and $(\ell=3)$ modes simultaneously, we can reject this
false match. The results of
these matching studies are summarized in Tables~\ref{tab:G2.5match}
and~\ref{tab:G3.5match}. As seen in the tables, the matching of the 3.5PN
and numerical waveforms are significantly better than the matching of
the 2.5PN and numerical waveforms for all modes. Similarly, all PN
modes match the numerical waveforms better over the shorter
integration times. This is
consistent with the qualitative agreements in the waveforms seen in
Figs.~\ref{fig:G2.5_re_h22}~--~\ref{fig:G3.5_re_h33}. Note that the
3.5PN and numerical waveform matches  for all modes 
are significantly better for the G3.5
configuration than the G2.5 configuration for the longer $t=1000M$ integration
time (the differences between the matches are most striking for
$(\ell=3,m=\pm3)$ modes, where the matching is $\sim0.7$ for G3.5 and
$\sim0.4$ for G2.5).
The only place where the matches for the G2.5 configuration  are consistently
better than the matches for the G3.5 configuration is the
is the $(\ell=2,m=\pm1)$ modes for the shorter integration
times. 
Thus, it appears that the 3.5PN waveforms, in general, produce
superior results for the more circular G3.5 configuration,
which is likely due to the fact that the higher PN order B waveforms 
are accurate for quasi-circular, rather than eccentric, binaries.
\begin{table} \caption{The overlap (matching) of the real and
imaginary parts  of the
modes of $h$
of the G2.5 configuration for the truncated 2.5PN and 3.5 PN
waveforms and the numerical waveforms for various integration times
and PN time translation $\delta t$. In all cases, we start the
integration just after the numerical initial data (spurious radiation)
pulse leaves the system.}
  \label{tab:G2.5match}
  \begin{ruledtabular}
  \begin{tabular}{lccc}
   Integration Time  & 600 & 800 & 1000 \\
\hline
Truncated 2.5PN $(\delta t = 112.2)$\\
\hline
  Re $(\ell=2, m=2)$ & 0.789 & 0.615 & 0.365 \\
  Re $(\ell=2, m=1)$ & 0.705 & 0.501 & 0.292 \\
  Re $(\ell=3, m=3)$ & 0.596 & 0.286 &-0.038 \\
\hline
\hline
3.5PN $(\delta t = 112.2)$\\
\hline
  Re $(\ell=2, m=2)$ & 0.975 & 0.922 & 0.693 \\
  Im $(\ell=2, m=2)$ & 0.976 & 0.924 & 0.723 \\
  Re $(\ell=2, m=-2)$ & 0.975 & 0.922 & 0.693 \\
  Im $(\ell=2, m=-2)$ & 0.978 & 0.926 & 0.723 \\

  Re $(\ell=2, m=1)$ & 0.982 & 0.938 & 0.687 \\
  Im $(\ell=2, m=1)$ & 0.977 & 0.924 & 0.699 \\
  Re $(\ell=2, m=-1)$ & 0.984 & 0.939 & 0.707 \\
  Im $(\ell=2, m=-1)$ & 0.980 & 0.933 & 0.711 \\

  Re $(\ell=3, m=3)$ & 0.908 & 0.794 & 0.418 \\
  Im $(\ell=3, m=3)$ & 0.916 & 0.795 & 0.435 \\
  Re $(\ell=3, m=-3)$ & 0.909 & 0.782 & 0.403 \\
  Im $(\ell=3, m=-3)$ & 0.912 & 0.794 & 0.426 \\
\hline
\hline
3.5PN $(\delta t = 233.3)$\\
\hline
  Re $(\ell=2, m=2)$ & 0.928 & 0.803 & 0.746 \\
  Re $(\ell=2, m=1)$ & 0.918 & 0.800 & 0.774 \\
  Re $(\ell=3, m=3)$ &-0.850 &-0.602 &-0.492 \\
  \end{tabular}
  \end{ruledtabular}
\end{table}
\begin{table}
  \caption{The overlap  of the real and imaginary parts of the modes of $h$
of the G3.5 configuration for the 3.5 PN waveforms and the
numerical waveforms. In all cases, we
start the integration just after the numerical 
initial data (junk radiation) pulse
leaves the system.}
  \label{tab:G3.5match}
  \begin{ruledtabular}
  \begin{tabular}{lccc}
   Integration Time  & 600 & 800 & 1000 \\
\hline
3.5PN $(\delta t = 112.5)$\\
\hline
  Re $(\ell=2, m=2)$ & 0.986 & 0.964 & 0.895 \\
  Im $(\ell=2, m=2)$ & 0.987 & 0.962 & 0.900 \\
  Re $(\ell=2, m=-2)$ & 0.986 & 0.964 & 0.895 \\
  Im $(\ell=2, m=-2)$ & 0.987 & 0.962 & 0.901 \\

  Re $(\ell=2, m=1)$ & 0.904 & 0.912 & 0.843 \\
  Im $(\ell=2, m=1)$ & 0.916 & 0.901 & 0.820 \\
  Re $(\ell=2, m=-1)$ & 0.920 & 0.908 & 0.833 \\
  Im $(\ell=2, m=-1)$ & 0.917 & 0.903 & 0.816 \\

  Re $(\ell=3, m=3)$ & 0.938 & 0.891 & 0.738 \\
  Im $(\ell=3, m=3)$ & 0.919 & 0.868 & 0.721 \\
  Re $(\ell=3, m=-3)$ & 0.931 & 0.880 & 0.733 \\
  Im $(\ell=3, m=-3)$ & 0.906 & 0.857 & 0.721 \\
  \end{tabular}
  \end{ruledtabular}
\end{table}

In Tables~\ref{tab:G2.5psi4match_altdt}-\ref{tab:G3.5psi4matchPd_altdt} we
show the matching of the modes of $\psi_4$ between 2.5PN, 3.5PN, and
the numerical $\psi_4$. Here we find a better match when we use a
slightly altered time offset. Note that matching is generally worse
than that observed with $h$, especially for the longer integration
times. This is consistent with the observation that the amplitude of
$\psi_4$ increases more rapidly in time than $h$ (due to the effects
of increasing frequency and the two time derivatives). Thus a matching
of $\psi_4$ emphasizes the disagreement between the PN and numerical
waveforms at later times. Interestingly, the matching in G3.5 is
significantly better than G2.5 for the $1000M$ integration time,
particularly in the $(\ell=3,m=3)$ mode, where the matching between
the 3.5PN and numerical $\psi_4$ is $65\%$ for G3.5 and only
$14\%$ for G2.5.
\begin{table} 
  \caption{The overlap (matching) of the real and imaginary parts  of the
modes of $\psi_4$
of the G2.5 configuration for the truncated 2.5PN and 3.5 PN
waveforms and the numerical waveforms for various integration times
with the PN time translation $\delta t = 106.5$ for the truncated 2.5PN
and $\delta t = 113.0$ for the 3.5PN. 
In all cases, we start the integration after t=180. 
The integration time means that 
the end of integration is the same as that used in the overlap of $h$.}
  \label{tab:G2.5psi4match_altdt}
  \begin{ruledtabular}
  \begin{tabular}{lccc}
   Integration Time  & 600 & 800 & 1000 \\
\hline
Truncated 2.5PN $(\delta t = 106.5)$\\
\hline
  Re $(\ell=2, m=2)$ & 0.900 & 0.744 & 0.435 \\
  Im $(\ell=2, m=2)$ & 0.898 & 0.717 & 0.469 \\

  Re $(\ell=2, m=1)$ & 0.824 & 0.654 & 0.408 \\
  Im $(\ell=2, m=1)$ & 0.851 & 0.675 & 0.431 \\

  Re $(\ell=3, m=3)$ & 0.767 & 0.472 & 0.00578 \\
  Im $(\ell=3, m=3)$ & 0.776 & 0.477 & 0.0102 \\
\hline
\hline
3.5PN $(\delta t = 113.0)$\\
\hline
  Re $(\ell=2, m=2)$ & 0.980 & 0.909 & 0.519 \\
  Im $(\ell=2, m=2)$ & 0.984 & 0.916 & 0.563 \\

  Re $(\ell=2, m=1)$ & 0.982 & 0.936 & 0.544 \\
  Im $(\ell=2, m=1)$ & 0.976 & 0.921 & 0.594 \\

  Re $(\ell=3, m=3)$ & 0.906 & 0.759 & 0.150 \\
  Im $(\ell=3, m=3)$ & 0.906 & 0.754 & 0.140 \\
  \end{tabular}
  \end{ruledtabular}
\end{table}
\begin{table}
  \caption{The overlap of the real and imaginary parts of the modes of
$\psi_4$ of the G3.5 configuration for the 3.5 PN waveforms and the
numerical waveforms with $\delta t = 113.5$.  In all cases, we start
the integration after t=180.  The integration time means that the end
of integration is the same as that used in the overlap of $h$.}
  \label{tab:G3.5psi4matchPd_altdt}
  \begin{ruledtabular}
  \begin{tabular}{lccc}
   Integration Time  & 600 & 800 & 1000 \\
\hline
3.5PN $(\delta t = 113.5)$\\
\hline
  Re $(\ell=2, m=2)$ & 0.981 & 0.962 & 0.860 \\
  Im $(\ell=2, m=2)$ & 0.983 & 0.958 & 0.876 \\

  Re $(\ell=2, m=1)$ & 0.882 & 0.927 & 0.850 \\
  Im $(\ell=2, m=1)$ & 0.853 & 0.893 & 0.811 \\

  Re $(\ell=3, m=3)$ & 0.869 & 0.841 & 0.640 \\
  Im $(\ell=3, m=3)$ & 0.868 & 0.834 & 0.649 \\
  \end{tabular}
  \end{ruledtabular}
\end{table}

\section{Conclusion}
\label{sec:discussion}

We analyzed the first long-term generic waveform produced by the
merger of unequal mass, unequal spins, precessing black-hole binaries (a
shorter
simulation of this kind, which led to the discovery of the very large
recoil configuration, was reported in~\cite{Campanelli:2007ew}).  We
demonstrated eighth-order convergence of the waveform phase and
fourth-order convergence of the amplitude (consistent with the order
of accuracy of the extraction routine) in the numerical results.
These waveforms clearly show the
effects of eccentricity and precession on the amplitude in the sub-leading
$(\ell=2,m=1)$  and $(\ell=3,m=3)$ modes. In particular, analyzing the
$(\ell=2,m=1)$ mode 
provides a way of detecting
precessional effects in the observed waveforms. We have also found
that there are two sources of eccentricity for a generic binary.
Residual eccentricity, due to a non-ideal choice of initial data
parameters that tends to damp out as the binary separation decreases,
and precession-induced eccentricity that grows as the orbital
separation falls below $\sim 15M$ (this increase in eccentricity at
later times is apparent in the $(\ell=3,m=3)$ mode of $h$ in both the
PN and numerical waveforms).

We have compared these waveforms with the truncated 2.5 post-Newtonian
waveforms, as well as the waveforms with the non-spinning 3.0 PN
conservative and 3.5 PN radiative corrections.  We find a good initial
agreement of waveforms for the first six cycles, with overlaps of
over $97\%$ for the $(\ell=2, m=\pm2)$ modes, $90\%$-$98\%$ for the
$(\ell=2,m=\pm1)$, and over $90\%$ for the
$(\ell=3,m=\pm3)$ modes. This provides a natural way to match
numerical waveforms to the post-Newtonian ones with a time translation
(the same for all modes)
motivated by the physical location of the observer
(See Fig.~\ref{fig:G3.5_re_h22}, for
instance).  The agreement degrades as we approach the more dynamical
region of the late merger and plunge.  The disagreement begins in a
region where the numerical waveform is still very accurate. Thus it appears
that the disagreement is mainly due to errors introduced by truncating
the PN series. Hence the overlap should be improved significantly
by including 3.0 PN and higher-order conservative and radiative
corrections, including spin
 terms~\cite{Steinhoff:2007mb, Porto:2007tt, Porto:2008tb,
Porto:2008jj, Steinhoff:2008zr}.  

In fact, our results indicate that higher-order PN corrections to the orbital
motion may further increase the accuracy of the PN waveforms.
Although, the PN expansion has not yet been shown to converge, we do
find remarkably better agreement in $\psi_4$ between the PN and
numerical waveforms when moving from a 2.5PN EOM to a 3.5 EOM. 
This would appear to 
underscore the need for higher-order post-Newtonian 
calculations of both spin-orbit and spin-spin terms (especially in the
EOM).  
Spin effects first appear at 1.5PN order, producing
the spin-orbit hangup effect~\cite{Campanelli:2006uy, Dain:2008ck}. Other
spin effects, such as those due to precession, generate more subtle
effects in the waveforms, and require higher-order PN corrections to
accurately model (while subtle, these effects are also responsible for
the very large kicks seen in spinning binaries with the spins oriented
in the orbital plane).
 Our results seem to indicate that calculating these
higher-order correction may prove to be invaluable for generating
waveform templates from generic black-binary configurations.

\acknowledgments
We thank E. Berti, A. Buonanno, A. Gopakumar, and R. Porto for careful
review of the manuscript, and  we thank B. Krishnan for suggesting the
technique to calculate $h$ from $\psi_4$.  We thank the referees for
their many helpful suggestions and insights, which have greatly
improved the paper.  We gratefully acknowledge NSF for financial
support from grant PHY-0722315, PHY-0653303, PHY 0714388, and PHY
0722703; and NASA for financial support from grant NASA
07-ATFP07-0158.  Computational resources were provided by Lonestar
cluster at TACC and by NewHorizons at RIT.

\appendix
\section{Transformation of the $(\ell,m)$ modes of spin-weighted
fields under arbitrary rotations.}
\label{app:proof}

Here we consider the spin-weighted spherical harmonics in two
different angular coordinate systems, $(\theta,\phi)$ and $(\theta',
\phi')$, related to each other by a simple rotation. For convenience,
we will use $\Omega$ to denote the coordinates $(\theta,\phi)$ and
$d \Omega = \sin \theta d\theta\,d\phi$ to denote the volume element
on the unit sphere. To construct spin-weighted functions, we need to
define a null dyad $q^A$ on the unit sphere obeying $q^A\,q_A=0$ and
$q^A\,\bar q_A = 2$ (indices are raised and lowered with the unit
sphere metric). Here we will use
$q^A = \partial_\theta + i/\sin\theta\, \partial_\phi$
(see~\cite{Gomez:1996ge} for a review of the subject). 
Any two choices for the dyad $q^A$ and $q'^A$ can differ by at most
a phase factor, i.e.\ $q'^A = e^{i \chi} q^A$.
A spin-weight $s$ field $J$ transforms as $J\to J' = e^{i s \chi} J$
under this change in spin-basis. 
Of relevance here are the two dyads $q^A = \partial_\theta +
i/\sin\theta\, \partial_\phi$ and $q'^A = \partial_{\theta'} +
i/\sin{\theta'}\, \partial_{\phi'}$. 
The choice of $q^A$ fixes the $\eth$ operator on spin-weighted
fields.

The spin-weighted  spherical harmonics are constructed as
 follows~\cite{Goldberg:1967},
\begin{eqnarray}
\label{eq:sylm}
\SYLM{s}{\ell}{m}(\Omega) = \sqrt{\frac{(\ell-|s|)!}{(\ell+|s|)!}}
  \left\{ \begin{array}{ll}
                  (-1)^s \eth^s Y_{\ell m}(\Omega) & \mbox{if $s>0$ } \\
              \bar\eth^{|s|} Y_{\ell m}(\Omega) & \mbox{if $s<0$ }
   \end{array} \right.,
\end{eqnarray}
where 
\begin{eqnarray}
  \eth f = \partial_\theta f + \frac{i}{\sin{\theta}}
    \partial_\phi f - s f \cot{\theta} \nonumber\\
  \bar\eth f = \partial_\theta f - \frac{i}{\sin{\theta}}
    \partial_\phi f + s f \cot{\theta}, \label{eq:eth}
\end{eqnarray}
for a function $f$ of spin-weight $s$. In the $\Omega'$ coordinates
and the corresponding $q'^A$ spin basis.
Eqs.~(\ref{eq:sylm})~and~(\ref{eq:eth}) take on an identical form,
but with the $\Omega$ coordinates replaced with the $\Omega'$
coordinates.
Let $J$ be a spin-weighted $s$ field of arbitrary spin-weight that can
be decomposed into spin-weighted spherical harmonics. That is,
\begin{equation}
  J = \sum_{\ell=|s|}^{\infty}\sum_{m=-\ell}^{\ell} J_{\ell m}
\SYLM{s}{\ell}{m} (\Omega).
\end{equation}
We define a spin-zero potential $j$, such that
\begin{equation}
  j = \sum_{\ell=|s|}^{\infty}\sum_{m=-\ell}^{\ell} j_{\ell m} Y_{\ell
m} (\Omega),
\end{equation}
where
\begin{eqnarray}
  \label{eq:jlmJlm}
  j_{\ell m} = J_{\ell m} \sqrt{\frac{(\ell-|s|)!}{(\ell+|s|)!}}\, p,
\end{eqnarray}
and
$p=(-1)^s$ if $s>0$ and $p=1$ otherwise.
Hence
\begin{eqnarray}
  J = 
\left\{ \begin{array}{ll}
                  \eth^s j & \mbox{if $s>0$ } \\
              \bar\eth^{|s|} j & \mbox{if $s<0$ }
   \end{array} \right..
\end{eqnarray}
Under a change of spin basis, $J\to J' = e^{i s \chi} J$ but
$j\to j' = j$. Thus
\begin{eqnarray}
  J' = 
\left\{ \begin{array}{ll}
                  \eth'^s j & \mbox{if $s>0$ } \\
              \bar\eth'^{|s|} j & \mbox{if $s<0$ }
   \end{array} \right.,
\end{eqnarray}
and
\begin{eqnarray}
  J' = \sum_{\ell=|s|}^{\infty} \sum_{m=-\ell}^{\ell}
\sqrt{\frac{(\ell+|s|)!}{(\ell-|s|)!}}\,p\, j'_{\ell m}
\SYLMP{s}{\ell}{m}(\Omega'),
\end{eqnarray}
where  
\begin{equation}
  j'_{\ell m} = \int j\, \overline{Y'_{\ell m}(\Omega')} d \Omega'.
\end{equation}
Thus
\begin{eqnarray}
  J_{\ell m} = j_{\ell m}\, \sqrt{\frac{(\ell+|s|)!}{(\ell-|s|)!}}\,p,
\nonumber\\
  J'_{\ell m} = j'_{\ell m}\, \sqrt{\frac{(\ell+|s|)!}{(\ell-|s|)!}}\,p,
  \label{eq:Jlmjlm}
\end{eqnarray}
where
\begin{equation}
  J'_{\ell m} = \int J'\, \overline{\SYLMP{s}{\ell}{m}(\Omega')} d \Omega',
\end{equation}
and hence we can determine how the modes of $J$ mix under a rotation
of the coordinates by looking at the modes of $j$.

It was shown in~\cite{Goldberg:1967} that the relationship between the 
spherical harmonic modes $Y_{\ell m} (\Omega)$ and $Y'_{\ell m} (\Omega')$,
where the $\Omega'$ coordinates are obtained from the $\Omega$ coordinates
by a rotation described by the Euler angles $\alpha$, $\beta$, $\gamma$ 
in Sec.~\ref{sec:PN-intro}, is given by
\begin{equation}
  Y_{\ell m} (\Omega) = \sum_{m'=-\ell}^{\ell}
     e^{-i (m' \alpha+ m \gamma)}\, d^{\ell}_{m' m}(-\beta)\,
       Y'_{\ell m'}(\Omega'),
\end{equation}
and hence
\begin{eqnarray}
  j_{\ell m} &=& \int j\, \overline{Y_{\ell m}(\Omega)} d \Omega \nonumber \\
             &=& \int j\, \overline{Y_{\ell m}(\Omega)} d \Omega' \nonumber \\
             &=& \int j \sum_{m'=-\ell}^{\ell}
     e^{i (m' \alpha +m\gamma)}\,
       d^{\ell}_{m' m}(-\beta)\, \overline{Y'_{\ell m'}(\Omega')}
       d\Omega' \nonumber \\
             &=& \sum_{m'=-\ell}^{\ell} e^{i (m' \alpha+m \gamma)}\,
          d^{\ell}_{m' m} (-\beta)\, j'_{\ell m'}.
\end{eqnarray}
Finally, using Eq.~(\ref{eq:Jlmjlm}) we get
\begin{equation}
J_{\ell m} = \sum_{m'=-\ell}^{\ell} e^{i (m' \alpha + m \gamma)}\,
        d^{\ell}_{m' m} (-\beta)\, J'_{\ell m},
\end{equation}
which is independent of the spin-weight of $J$.

\bibliographystyle{apsrev}
\bibliography{../../Bibtex/references}

\end{document}